\documentclass[12pt,preprint]{aastex}

\slugcomment{Version: \today}

\usepackage{color}
\usepackage{graphicx}

\newcommand{\degree}{\ensuremath{^\circ}}
\def\ga{\mathrel{\raise0.35ex\hbox{$\scriptstyle >$}\kern-0.6em
\lower0.40ex\hbox{{$\scriptstyle \sim$}}}}
\def\la{\mathrel{\raise0.35ex\hbox{$\scriptstyle <$}\kern-0.6em
\lower0.40ex\hbox{{$\scriptstyle \sim$}}}}
\def\co{CO {J}=1--0 }
\def\cotwo{CO {J}=2--1 }

\def\hij{high-{J} }
\def\loj{low-{J} }
\def\ee #1 {\times 10^{#1}}          
\def\ut #1 #2 { \, \mathrm{#1}^{#2}} 
\def\u #1 { \, \mathrm{#1}}          
\def\msol{\hbox{$\hbox{M}_\odot$}}

\begin{document}

\title{The Science Cases for Building \\
a Band 1 Receiver Suite for ALMA}

\author{
J.~Di~Francesco\altaffilmark{1,2},
D.~Johnstone\altaffilmark{1,2},
B.~Matthews\altaffilmark{1,2},
N.~Bartel\altaffilmark{3},
L.~Bronfman\altaffilmark{4},
S.~Casassus\altaffilmark{4},
S.~Chitsazzadeh\altaffilmark{2,5},
M.~Cunningham\altaffilmark{6},
G.~Duch\^{e}ne\altaffilmark{7,8},
J.~Geisbuesch\altaffilmark{9},
A.~Hales\altaffilmark{10},
P.T.P.~Ho\altaffilmark{11}
M.~Houde\altaffilmark{5},
D.~Iono\altaffilmark{12},
F.~Kemper\altaffilmark{11},
P.M.~Koch\altaffilmark{11},
K.~Kohno\altaffilmark{13},
R.~Kothes\altaffilmark{9},
S-P.~Lai\altaffilmark{14},
K.Y.~Lin\altaffilmark{11},
S.-Y.~Liu\altaffilmark{11},
B.~Mason\altaffilmark{10},
T.J.~Maccarone\altaffilmark{15},
N.~Mizuno\altaffilmark{12},
O.~Morata\altaffilmark{11},
G.~Schieven\altaffilmark{1},
A.M.M.~Scaife\altaffilmark{15},
D.~Scott\altaffilmark{16},
H.~Shang\altaffilmark{11},
S.~Takakuwa\altaffilmark{11},
J.~Wagg\altaffilmark{17,18},
A.~Wootten\altaffilmark{10},
F.~Yusef-Zadeh\altaffilmark{19}
}

\altaffiltext{1}{National Research Council Canada, 5071 West Saanich Rd, Victoria, BC, V9E 2E7, Canada}
\altaffiltext{2}{Dept.\ of Physics \& Astronomy, University of Victoria, Victoria, BC, V8P 1A1, Canada}
\altaffiltext{3}{Dept.\ of Physics and Astronomy, York University, Toronto, M3J 1P3, ON, Canada}
\altaffiltext{4}{Dept.\ de Astronom\'{\i}a, Universidad de Chile, Casilla 36-D, Santiago, Chile}
\altaffiltext{5}{Dept.\ of Physics and Astronomy, The University of Western Ontario, London, ON, N6A 3K7, Canada}
\altaffiltext{6}{School of Physics, University of New South Wales, Sydney, NSW 20152, Australia}
\altaffiltext{7}{Astronomy Dept., University of California, Berkeley, CA 94720-3411, USA}
\altaffiltext{8}{Universit\'e Joseph Fourier - Grenoble 1/CNRS, LAOG UMR 5571, BP 53, 38041 Grenoble, France}
\altaffiltext{9}{National Research Council Canada, P.O.\ Box 248, Penticton, BC, V2A 6J9, Canada}
\altaffiltext{10}{National Radio Astronomy Observatory, 520 Edgemont Road, Charlottesville, Virginia 22903, USA}
\altaffiltext{11}{Academia Sinica, Institute of Astronomy and Astrophysics, P.O.\ Box 23-141, Taipei 10617, Taiwan}
\altaffiltext{12}{National Astronomical Observatory of Japan, 2-21-1 Osawa, Mitaka, Tokyo, 181-8588, Japan}
\altaffiltext{13}{Institute of Astronomy, The University of Tokyo, 2-21-1 Osawa, Mitaka,Tokyo 181-0015, Japan}
\altaffiltext{14}{Institute of Astronomy and Dept.\ of Physics, National Tsing Hua University, Taiwan}
\altaffiltext{15}{School of Physics and Astronomy, University of Southampton, Southampton, Hampshire, S017 1BJ, UK}
\altaffiltext{16}{Dept.\ of Physics and Astronomy, University of British Columbia, Vancouver, BC, V6T 1Z1, Canada}
\altaffiltext{17}{European Southern Observatory, Alonso de Cordova 3107, Vitacura, Casilla 19001, Santiago 19, Chile}
\altaffiltext{18}{Astrophysics Group, Cavendish Laboratory, JJ Thomson Avenue, Cambridge, CB30HE, UK}
\altaffiltext{19}{Dept.\ of Physics and Astronomy and Center for Interdisciplinary Research in Astronomy, Northwestern 
University, Evanston, IL 60208, USA}

~
\section{Executive Summary}
\label{sec:exec}

We present a set of compelling science cases for the ALMA Band 1 receiver suite.  For these cases, we assume in
tandem the updated nominal Band 1 frequency range of 35-50 GHz with a likely extension up to 52 GHz; together these 
frequencies optimize the Band 1 science return.  The scope of the science cases ranges from nearby stars to the re-
ionization edge of the Universe.  Two cases provide additional leverage on the present ALMA Level One Science Goals 
and are seen as particularly powerful motivations for building the Band 1 Receiver suite: (1) detailing the evolution of grains 
in protoplanetary disks, as a complement to the gas kinematics, requires continuum observations out to $\sim 35\,$GHz ($
\sim 9\,$mm); and (2) detecting CO 3--2 line emission from galaxies like the Milky Way during the epoch of re-ionization, 
i.e., 6 $< z <$ 10, also requires Band 1 receiver coverage.  The range of Band 1 science is wide, however, and includes
studies of very small dust grains in the ISM, pulsar wind nebulae, radio supernovae, X-ray binaries, the Galactic Center 
(i.e., Sgr A*), dense cloud cores, complex carbon-chain molecules, masers, magnetic fields in the dense ISM, jets and outflows from young stars, distant galaxies, and galaxy clusters (i.e., the Sunyaev-Zel'dovich Effect).  A comparison of ALMA and the Jansky VLA (JVLA) at the same frequencies of Band 1 finds similar sensitivity performance at 40--50 GHz, with a slight edge for ALMA at higher frequencies (e.g.,., within a factor of $2$ for continuum observations).  With its larger number of instantaneous baselines, however, ALMA Band 1data will have greater fidelity than those from the JVLA at similar 
frequencies.
\newpage

\section{Introduction}
\label{sec:intro}

The Atacama Large Millimeter/submillimeter Array (ALMA) will be a single research instrument composed of at least fifty-four 12-m and twelve 7-m high-precision antennas, located at a very high altitude of 5000 m on the Chajnantor plain of the Chilean Andes.  The weather conditions at the ALMA site will allow transformational research into the physics of the cold Universe across a wide range of wavelengths, from radio to submillimeter.  Thus, ALMA will be capable of probing the first stars and galaxies and directly imaging the disks in which planets are formed. ALMA will be a complete astronomical imaging and spectroscopic instrument for the millimeter/submillimeter regime, providing scientists with capabilities and wavelength coverage that complement those of other research facilities of its era, such as the Jansky Very Large Array (JVLA), James Webb Space Telescope (JWST), 30-m class Giant Segmented Mirror Telescopes (GSMTs), and the Square Kilometer Array (SKA).  ALMA will be the pre-eminent platform for astronomical research at millimeter and submillimeter wavelengths for decades to come.

ALMA will revolutionize many areas of astronomy and an amazing breadth of science has already been proposed (see for example the ALMA Design Reference Science Plan). The technical requirements of the ALMA Project are, however, driven by three specific Level One Science Goals:

\noindent{\bf (1)} The ability to detect spectral line emission from CO or CII in a normal galaxy like the Milky Way at a redshift of $z = 3$, in less than 24 hours of observation.\\
\noindent{\bf (2)} The ability to image the gas kinematics in a solar-mass protostellar/ protoplanetary disk at a distance of 150 pc (roughly, the distance of the star-forming clouds in Ophiuchus or Corona Australis), enabling one to study the physical, chemical, and magnetic field structure of the disk and to detect the tidal gaps created by planets undergoing formation.\\
\noindent{\bf (3)} The ability to provide precise images at an angular resolution of 0.1$^{\prime\prime}$. Here the term ``precise image" means an accurate representation of the sky brightness at all points where the brightness is greater than 0.1\% of the peak image brightness. This requirement applies to all sources visible to ALMA that transit at an elevation greater than 20\degree.

ALMA was originally envisioned to provide access to all frequencies between 31 GHz and 950 GHz  accessible from the ground.  During a re-baselining exercise undertaken in 2001, the entire project was scrutinized to find necessary cost savings.  The two lowest receiver frequencies, Bands 1 and 2, covering 31--45 GHz and 67--90 GHz respectively, were among those items delayed beyond the start of science operations. Nevertheless, Band 1 was re-affirmed as a high priority future item for ALMA.

In May 2001, John Richer and Geoff Blake prepared a document {\it Science with Band 1 (31--45 GHz) on ALMA} as part of the re-baselining exercise. Key arguments for maintaining the receiver were put forward and included: (1) exciting science opportunities, bringing in a wider community of users; (2) a significantly faster imaging and survey instrument than the upgraded VLA (now known as the Jansky VLA or JVLA), especially due to the larger primary beam; (3) access to the southern sky at these wavelengths; (4) excellent science possible even in ``poor" weather; (5) a relatively cheap and reliable receiver to build and maintain; and (6) a very useful engineering/debugging tool for the entire array given the lower contribution of the atmosphere at many of its frequencies relative to other bands.

The Richer/Blake document was followed by an ASAC Committee Report in October 2001 in which the addition of Japan into the ALMA project re-opened the question of observing frequency priorities for those receiver bands which had been put on hold during re-baselining. The unanimous recommendation of the ASAC was to put Band 10 as top priority, followed by a very high priority Band 1. The key science cases for Band 1, at that time, were seen to be (1) high-resolution Sunyaev-Zel'dovich effect (SZE) imaging of cluster gas at all redshifts; and (2) mapping the cold ISM in Galaxies at intermediate and high redshift.

The scientific landscape has changed significantly since 2001 and thus it is time to update the main science 
drivers for Band 1, even reconsidering its nominal  frequency range to optimize its science return.  In addition, 
the ALMA Development process has begun, and now is the time to put forth the best case for longer wavelength 
observing with ALMA.  In October 2008, two dozen astronomers from around the globe met in Victoria, Canada 
to discuss Band 1 science.  This paper represents a brief summary of the outstanding cases made possible with 
Band 1 that were highlighted at that meeting and since.  In Section \ref{sec:freq}, we describe the new nominal 
Band 1 frequency range of 35-50 GHz, and its possible extension to 52 GHz.  In Section \ref{sec:one}, we 
present two science cases that reaffirm and enhance the already established ALMA Project Level One Science 
Goals.  In Section \ref{sec:range}, we provide a selection of science cases that reinforce the breadth and 
versatility of the Band 1 Receiver.  Section \ref{sec:consid} discusses both weather considerations at the ALMA 
site and the complementarity of ALMA versus the EVLA at Band 1 frequencies. Finally, Section \ref{sec:conc} 
briefly summarizes the report.

\section{The Band 1 Frequency Range}
\label{sec:freq}

Band 1 was originally defined as 31.3--45 GHz.  The lower value was the set at the lower edge of a frequency range assigned to radio astronomy and the upper value was set to include SiO $J$=1--0 emission at 43 GHz. Receiver technology advances, however, have made it possible to widen and shift the Band 1 range and optimize science return.  For example, a shift to higher frequencies for Band 1 will improve (slightly) the angular resolution of continuum observations and better exploit the advantages of the dry ALMA site.  Furthermore, a wider range and shift to higher frequencies will allow molecular emission from galaxies at a wider range of (slightly lower) redshifts to be explored, and allow molecular emission from several new species in our Galaxy to be probed.  (Of course, this shift does in turn remove the ability to detect molecular emission from some higher redshift galaxies and some other Galactic transitions.)

A review of the nominal frequency range by the Band 1 Science Team (i.e., several authors of this document) in June 2012 resulted in a proposed new Band 1 frequency range definition, nominally 35--50 GHz with an extension up to 52 GHz encouraged.  The shift up to 50 GHz will allow the important line CS $J$=1--0 at 48.99 GHz to be observable with ALMA. In addition, the nominal range of 35-50 GHz alone is itself $\sim$10\% wider than before.  As it will provide the highest sensitivities, the nominal range will be preferred for high-redshift science.  The extension to 50--52 GHz, which the JVLA cannot observe, may be somewhat adversely affected by atmospheric O$_{2}$, resulting in lower relative sensitivity.  Since numerous transitions of other interesting molecules have rest frequencies at 50-52 GHz, however, this extension will allow such emission to be probed toward sources in our Galaxy.  This document has been updated in September 2012 to reflect the new nominal frequency range and the extension.  See Section~\ref{sec:consid} for a comparison of the sensitivities and imaging characteristics of ALMA and the JVLA over their frequencies in common.

\section{Level One Science Cases for Band 1}
\label{sec:one}

In this section, we present two science cases that reaffirm and enhance the already established ALMA Project Level One Science Goals: Evolution of Grains in Disks Around Stars (\ref{sec:disk}) and The First Generation of Galaxies (\ref{sec:red}).

\subsection{Evolution of Grains in Disks Around Stars}
\label{sec:disk}
\subsubsection{Protoplanetary Disks}

Planet formation takes place in disks of dust and gas surrounding
young stars. It is within these gas-rich protoplanetary disks that
dust grains must agglomerate from the sub-micron sizes associated with
the interstellar medium to larger pebbles, rocks and planetesimals, if
planets are ultimately to be formed.
The timescale of this agglomeration process is
thought to be a few tens of Myr for terrestrial planets, while the
process leading to the formation of giant planet cores remains
uncertain. Core accretion models require at least a few Myr to form
Jovian planets (Pollack et al.\ 1996), while dynamical instability
models could form giant planets on orbital timescales ($t \ll 1$ Myr;
Boss 2005).

\begin{figure}
\begin{center}
\includegraphics[scale=0.65]{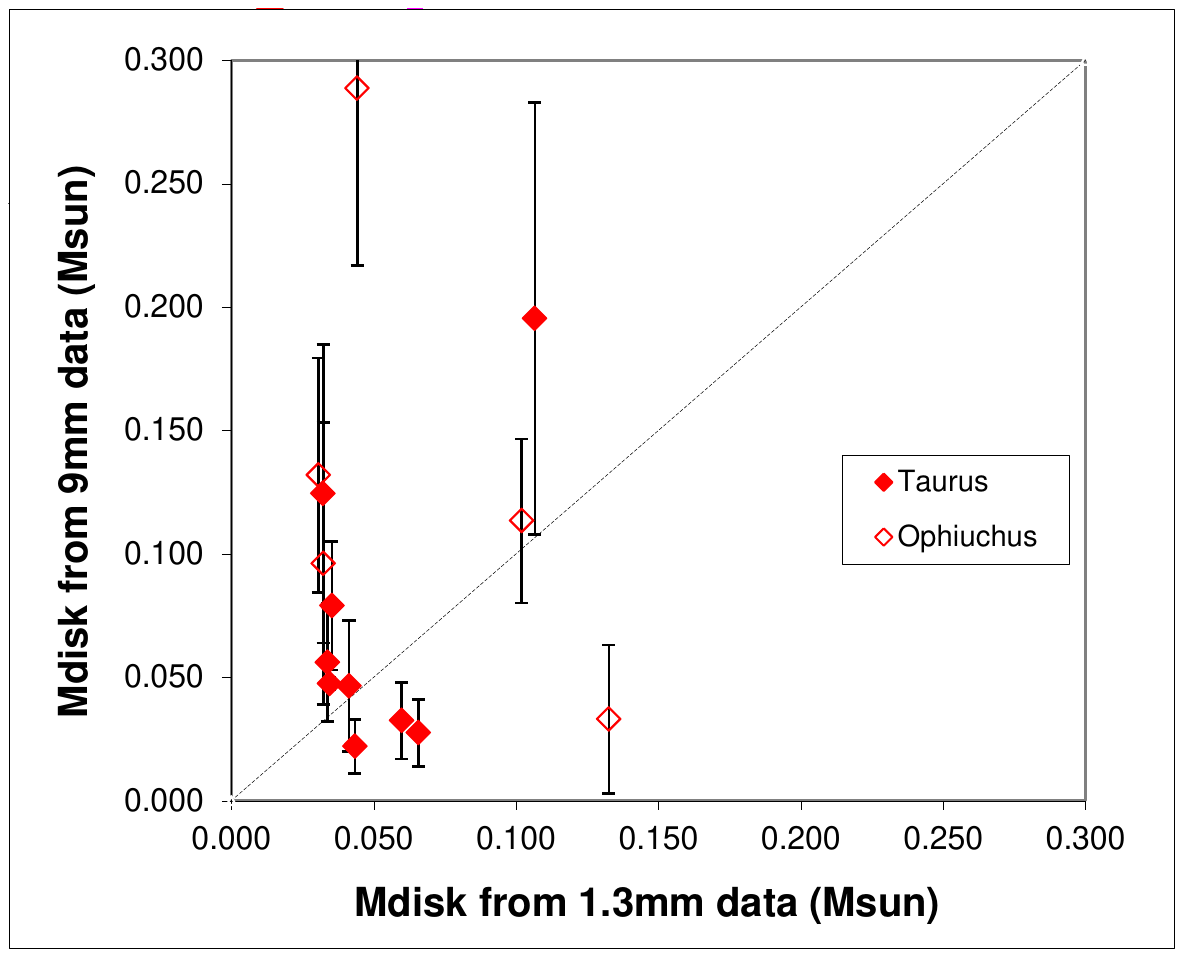}
\caption{Disk masses measured from 9 mm continuum emission compared to those measured from 1.3 mm continuum emission in the regions of Taurus and Ophiuchus.  Many disks show higher mass measurements at the longer wavelength, indicating the presence of larger grains than those detected at 1.3 mm measurements. (Greaves et al., in prep.)}
\label{fig:diskmasses}
\end{center}
\end{figure}

Gravitational instability models require high disk masses in order to
form planets. So far, most accurate disk mass estimates come from
 submillimeter and mm
observations, where the dust is optically thin. Andrews
\& Williams (2007a, 2007b) show that submillimeter
observations of dozens of protoplanetary disks reveal that only
one system could be gravitationally unstable, conflicting with the
high frequency of Jovian planets seen around low mass stars. Have
these relatively young (1--6 Myr) systems already formed planets, or is
most of the dust mass locked into larger grains and therefore not
accounted for in  submillimeter and millimeter observations?  If grain growth to
centimeter sizes has occurred, most of a disk's dust mass would reside in
the large particle population, which would emit at longer millimeter and 
centimeter wavelengths. Figure \ref{fig:diskmasses} from Greaves et al.\ (in prep.) compares disk masses for
objects in Taurus and Ophiuchus derived from 9 mm and 1.3 mm  dust
fluxes. The longer wavelength masses are found to generally be higher than the shorter wavelength values, indicating that a significant
fraction of the disks' total dust masses are indeed locked up in
larger grains.

Identifying {\it where} and {\it when} dust coagulation occurs are
critical to constrain current models of planetary formation. Growth
from sub-micron to micron-sized particles can be traced with infrared
spectroscopy and imaging polarimetry. The next step, growth beyond 
micron sizes, is readily studied by determining the slope of the
spectral energy distribution (SED) of the dust thermal emission at
 submillimeter and millimeter wavelengths. The dust mass opacity index at wavelengths 
longer than 0.1 millimeter is
approximately a power-law whose normalization depends on the dust
properties, such as composition, size distribution, and geometry
(Draine 2006). The index of the power law is $\beta$.  The presence of
large grains is detectable through a decrease in the index $\beta$,
which can be derived directly from the slope of the Rayleigh-Jeans
tail of the SED, $\alpha$, where $\beta = \alpha -2$, when the emission is optically thin. 
Studies reveal that the $\beta$ values of disks are substantially
lower than the typical ISM value of $\sim 2$ (e.g., Testi et al.\ 2003; Weintraub et
al.\ 1989; Adams et al.\ 1990; Beckwith et al.\ 1990; Beckwith \& Sargent
1991; Mannings \& Emerson 1994).

The key stumbling block to the interpretation of $\beta$ occurs when
the disk is not resolved spatially.  The amount of flux detected at a
given wavelength is a function of both $\beta$ and the size of the
disk (Testi et al.\ 2001). Resolving the ambiguity therefore is truly a
matter of resolution, and sufficient resolution is only offered at
these wavelengths by interferometers.

Among the three high level science goals of ALMA is the ability to
detect and image gas kinematics in protoplanetary disks undergoing
planetary formation at 150~pc.  At ALMA's observing wavelengths, its
capability for imaging the continuum dust emission in these disks is
also second-to-none.  At present, however, ALMA will only reach a
longest wavelength of 3.6 mm. Given that dust particles emit very
inefficiently at wavelengths longer than their sizes, {\it the present
ALMA design will not be sensitive to particles larger than $\sim 3$ mm.
This situation negates ALMA's potential ability to follow the dust grain
growth from mm-sized to cm-sized pebbles in protoplanetary disks.}

\begin{figure}
\begin{center}
\includegraphics[scale=0.65]{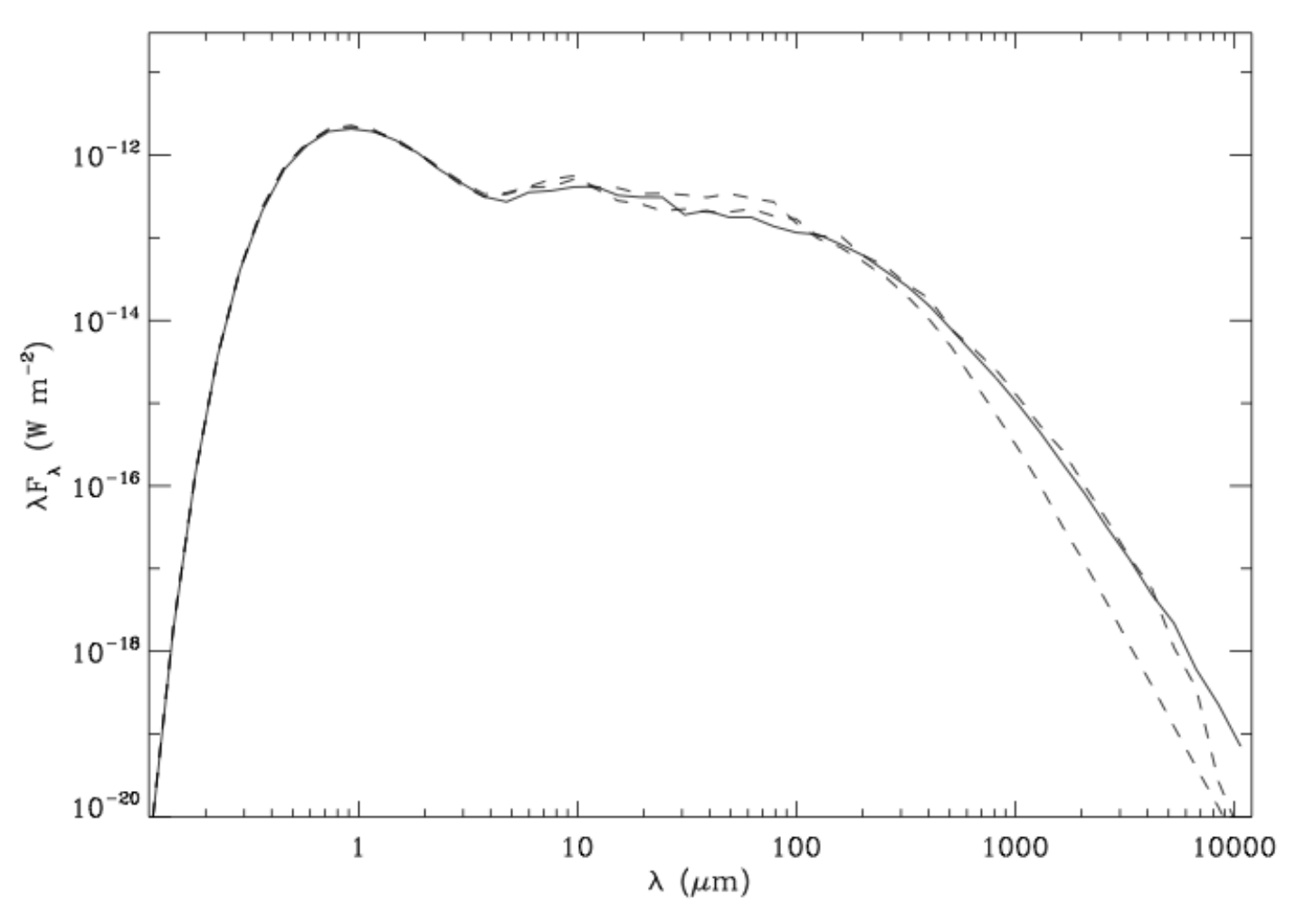}
\caption{Spectral energy distribution plot showing the differences
between three disk models having different maximum grain sizes.  The
solid curve is the model with a$_{\rm max}=1 $ cm, which keeps declining with
roughly constant slope all the way to 1 cm. The two dashed curves are
for a$_{\rm max}=10\ \mu$m and 1 mm. The top one, which breaks around 5 mm is the
model with a$_{\rm max}=1 $ mm. It's interesting to note how the fluxes are very
much the same for a$_{\rm max}= 1 $ mm or 1 cm, except precisely towards ALMA's
Band 1.  There is at least an order of magnitude difference in power
at 1 cm between the max$_{\rm size}=1 $ mm versus the max$_{\rm size}=1 $ cm disks. These
models indicate that observations at the ALMA Band 1 regime are crucial
for determining whether grain-growth to cm-sizes is indeed occurring.}
\label{fig:disk1}
\end{center}
\end{figure}

Figure~\ref{fig:disk1} shows the SEDs for three different circumstellar disk
models, computed using the full dust radiative transfer MCFOST code
(Pinte et al.\ 2006; Pinte et al.\ 2009). The model parameters are
representative of protoplanetary disks (although there is substantial
object-to-object variation). The circumstellar disk is passively
heated by a 4000 K, 2 L$_{\odot}$ central star and the system is located
160 pc away. The dust component of the disk is assumed to be fully
mixed with the gas and the latter is assumed to be in vertical
hydrostatic equilibrium. The disk extends radially from 1 AU to
100 AU. The total dust mass in the model is $10^{-3}$ M$_\odot$ (the gas
component is irrelevant for continuum emission calculations, so its
mass is not set in the model, though a typical 100:1 gas:dust ratio is
generally assumed). The dust population is described by a single power-law
size distribution $N(a)\propto a^{-3.5}$ with a minimum grain size of 0.03
$\mu$m and extending to 10 $\mu$m, 1 mm or 1 cm depending on the
model. The dust composition is taken to be the ``astronomical
silicates" from Draine (2003).

Figure~\ref{fig:disk1} reveals that observations in the ALMA Band 1 spectral
region are crucial for determining whether grain-growth to cm-sizes is
indeed occurring. The 1 cm flux density of the max$_{size}=$1\,cm disk model is
$\sim50\,\mu$Jy, comparable to the 1-sigma sensitivities provided by
ALMA's Band 1 with 1 minute integration. Besides ALMA, there are no
existing or planned southern astronomical facilities capable of
observing to such depths at these frequencies. Therefore, {\it if Band
1 is not built there will be no way of putting ALMA observations of
protoplanetary disks in the context of coagulation of
dust grains to centimeter sizes}.

\begin{figure}
\begin{center}
\includegraphics[scale=0.65]{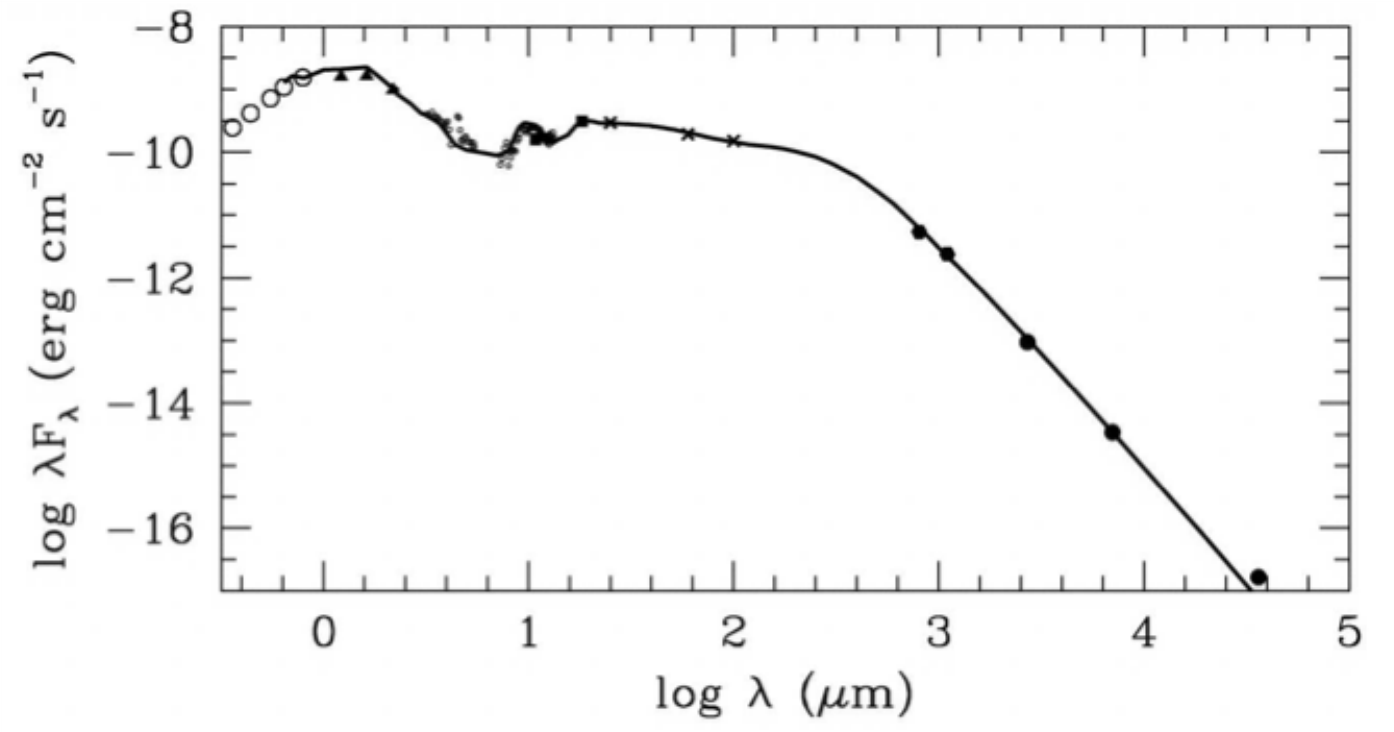}
\caption{Spectral energy distribution of TW Hya, showing
the fit to the SED for an irradiated accretion disk model with a maximum
particle size of 1 cm (Calvet et al.\ 2002).}
\label{fig:disk2}
\end{center}
\end{figure}

By complementing observations in other ALMA Bands, Band 1 will provide a
crucial longer wavelength lever to minimize the uncertainty in
$\alpha$.  Evidence for small pebbles has
been detected in several disks (Rodmann et al.\ 2006). The prime
example is TW Hya, a protoplanetary disk 50 pc from the Sun (Wilner et
al.\ 2000).  Its SED is well matched by an irradiated accretion disk
model fit from 10s of AU to an outer radius of 200 AU and requires the
presence of particle sizes up to 1 cm in the disk (see Figure~\ref{fig:disk2}). 
The measured $\beta$ is $0.7 \pm 0.1$ (Calvet et al.\ 2002).  To date,
no trend in $\beta$ has been detected with stellar luminosity, mass or
age (Ricci et al. 2010). Lower $\alpha$ values are associated with less
60 $\mu$m excess,
however, suggesting that settling or agglomeration processes could be
removing the smallest grains, decreasing the shorter wavelength
emission (Acke et al.\ 2004).

At the resolution provided by its longest baselines at $\sim$40 GHz
($\sim$0.14$^{\prime\prime}$), ALMA will easily resolve protoplanetary disks
at the distance of the closest star-forming regions (50--150 pc).  These
resolved images will provide the most accurate determination of the
disk's dust mass.  The dust distribution at centimeter wavelengths can then be
compared to millimeter and submillimeter images, revealing where in the disk 
dust coagulation is occurring.  For example, previous investigations of the
radial dependency of dust properties in disks by Guilloteau et al.\ (2009) and
Isella et al.\ (2010) were conducted at 1 mm and 3 mm, and as such they were
sensitive to only millimeter-sized grains.  Note, however, that Melis et al.\ 
(2011) used the Jansky VLA to map the 7 mm emission from the protoplanetary
disk around the young source L1527 IRS at $\sim$1.5$^{\prime\prime}$ and
tentatively detected a {\it dearth}\/ of ``pebble-sized" grains. ALMA Band 1
will help clarify this situation.

\begin{figure}
\begin{center}
\includegraphics[scale=0.65]{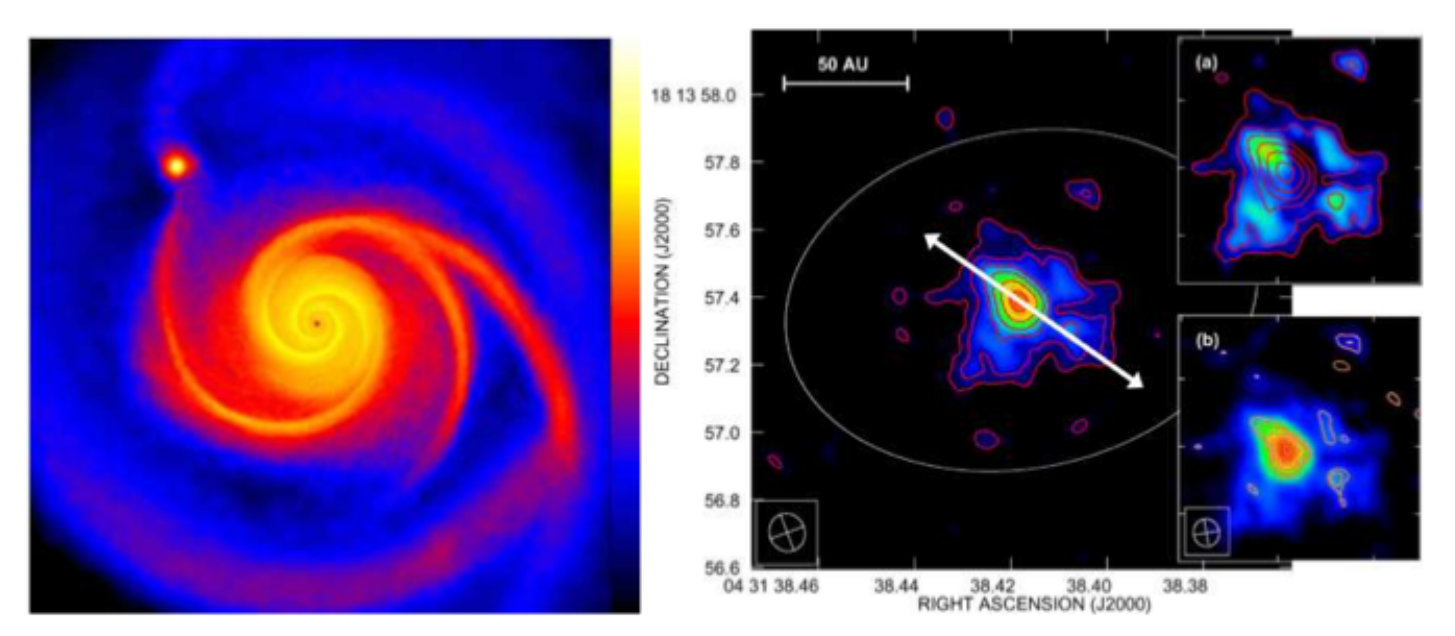}
\caption{(Left) Image from an SPH simulation showing the surface
density structure of a 0.3~M$_\odot$ disk around a 0.5~M$_\odot$ star. A
single dense clump has formed in the disk, at a radius of 75 AU and
with a mass of $\sim 8$ M$_{\rm Jup}$. {(Right)} VLA 1.3 cm images toward HL
Tau. The main image shows natural weighting with a beam of
0.11$^{\prime\prime}$ FWHM. The arrow indicates the jet axis. Upper inset:
compact central peak subtracted. Lower inset: uniform weighting, with a
beam of 0.08$^{\prime\prime}$ FWHM.  The compact object lies to the upper
right hand side. This condensation was also detected at 1.4 mm with the
BIMA array (Welch et al.\ 2004).}
\label{fig:disk3}
\end{center}
\end{figure}

As described above, Band 1 data will be sensitive to larger grains.  Moreover,
through detection of concentrations of such large grains, protoplanets in formation
can be identified. These condensations are predicted by simulations of gravitational
instability models (see Figure~\ref{fig:disk3}a; Greaves et al.\ 2008) and have been
detected in the nearby star HL Tau (Figure~\ref{fig:disk3}b; Greaves et al.\ 2008).
ALMA's high resolution is clearly critical to the detection of such substructures within
protoplanetary disks. 

Detecting dust emission at centimeter wavelengths also requires high sensitivity,
because its brightness is several orders of magnitude lower than in the
submillimeter.  In addition, at wavelengths longer than 7 mm (i.e., $\nu$ $<$ 
45 GHz), the contribution
from other radiative processes, such as ionized winds, can contribute
significantly to the total flux and complicate the interpretation of
detected emission. 
Rodmann et al.\ (2006) found that the contribution of free-free
emission to the total flux is typically 25\% at a wavelength of 7
mm.  Observations of continuum emission at the 35-50 GHz
(9 mm--6 mm) spectral
range enabled by Band 1 would increase substantially the sampling
rate in the region where emission is detected from both the free-free
and thermal dust emission components.  
The synergy with the JVLA will provide a longer wavelength lever for sources
observed in common, providing an estimate of the free-free contribution to the
Band 1 flux.  Such data would not be essential, however, given wide frequency
coverage within Band 1 alone.  For example, multiple continuum observations
could be used to quantify accurately the relative amounts of free-free and dust
emission through changes in spectral slope, and thereby determine precisely
the contribution from large dust grains (i.e., protoplanetary material).

\subsubsection{Debris Disks}

Around main sequence stars, pebble-sized bodies are produced differently 
than in disks around pre-main-sequence stars.  Here, destructive, collisional
cascades from even larger planetesimals through to centimeter, millimeter,
and then micron-sized particles provides ongoing replenishment of the debris
population (Wyatt 2009; Dullemond \& Dominik 2005).  The methods
of detection of large (i.e., centimeter-sized) grains is the same as in
protoplanetary disks, despite their origin in destructive, rather than
agglomerative, processes. In each case, the longer the wavelength at 
which continuum emission is detected, the larger the grains that must be 
present in the system.

Boley et al.\ (2012) detected the debris disk of Fomalhaut using Band 7,
and noted its sharp inner and outer boundary.  Band 1 images , however,
could show higher contrast features in debris disks compared to other ALMA
bands, due to the longer resonant lifetimes of
the larger particles that dominate the emission.  This sensitivity in turn will
help detect any edges and gaps in the disks. 
Dramatic changes in the morphology of debris disks as a function of wavelength 
have already been observed (e.g., Maness et al.\ 2008), but not yet at 
the long wavelengths Band 1 will probe. When observed,
such structures are often considered signposts to the existence of planets.  
These detections will be challenging compared with
detecting forming condensations in protoplanetary disks.  Debris disks 
typically have relatively low surface brightnesses and large spatial distributions
100s of AU in radii.  They also can be found much closer to the Sun than
protoplanetary disks.  Indeed, the closest disks could subtend as much as
$\sim$150$^{\prime\prime}$ on the sky (assuming a 300 AU diameter disk at 2 pc).
Therefore, ALMA's larger field of view than other long wavelength instruments,
such as the JVLA, will be very advantageous for imaging these objects, although
mosaicking will still be required to image the largest ones on the sky. ACA observations will
be instrumental in this research.

\subsection{The First Generation of Galaxies: \\ Molecular gas in galaxies during the epoch of re-ionization}
\label{sec:red}

The first generation of luminous objects in the Universe began the process of re-ionizing  the intergalactic 
medium (IGM).  The detection of large-scale polarization in the cosmic microwave background (CMB), caused 
by Thomson scattering of the CMB by the IGM during re-ionization, suggests that the Universe was significantly 
ionized as far back as $z \approx$ 11.0 $\pm$ 1.4 (Dunkley et al.~2009).  The ``near" edge of the era of re-
ionization has been inferred from the detection of the Gunn-Peterson effect (Gunn \& Peterson 1965) toward 
galaxies with $z \gtrsim 6$ (Fan et al.\ 2006a,b).  The nearly complete absorption of all continuum shortward of 
the Ly$\alpha$ break is due to moderate amounts of neutral hydrogen in the IGM, suggesting re-ionization was 
complete by $z$ $\approx$ 6.  The Gunn-Peterson effect also insures that at these redshifts the Universe is 
opaque at wavelengths shorter than $\sim$ 1$\,\mu$m.

To study the first generations of galaxies, and to understand the origins of the black hole-bulge mass relation, it will be necessary to study the star-formation properties of galaxies in the $6 \lesssim z \lesssim 11$ range.  Quasar hosts and other sources are rapidly being discovered at the near end of this range (e.g., Cool et al.~2006;  Mortlock et al.~2008; Glikman et al.~2008; Willott et al.~2009), and searches are underway for even more distant objects (e.g., Ota et al.~2008; Bouwens et al.~2009).

\begin{figure} 
\includegraphics[scale=0.65]{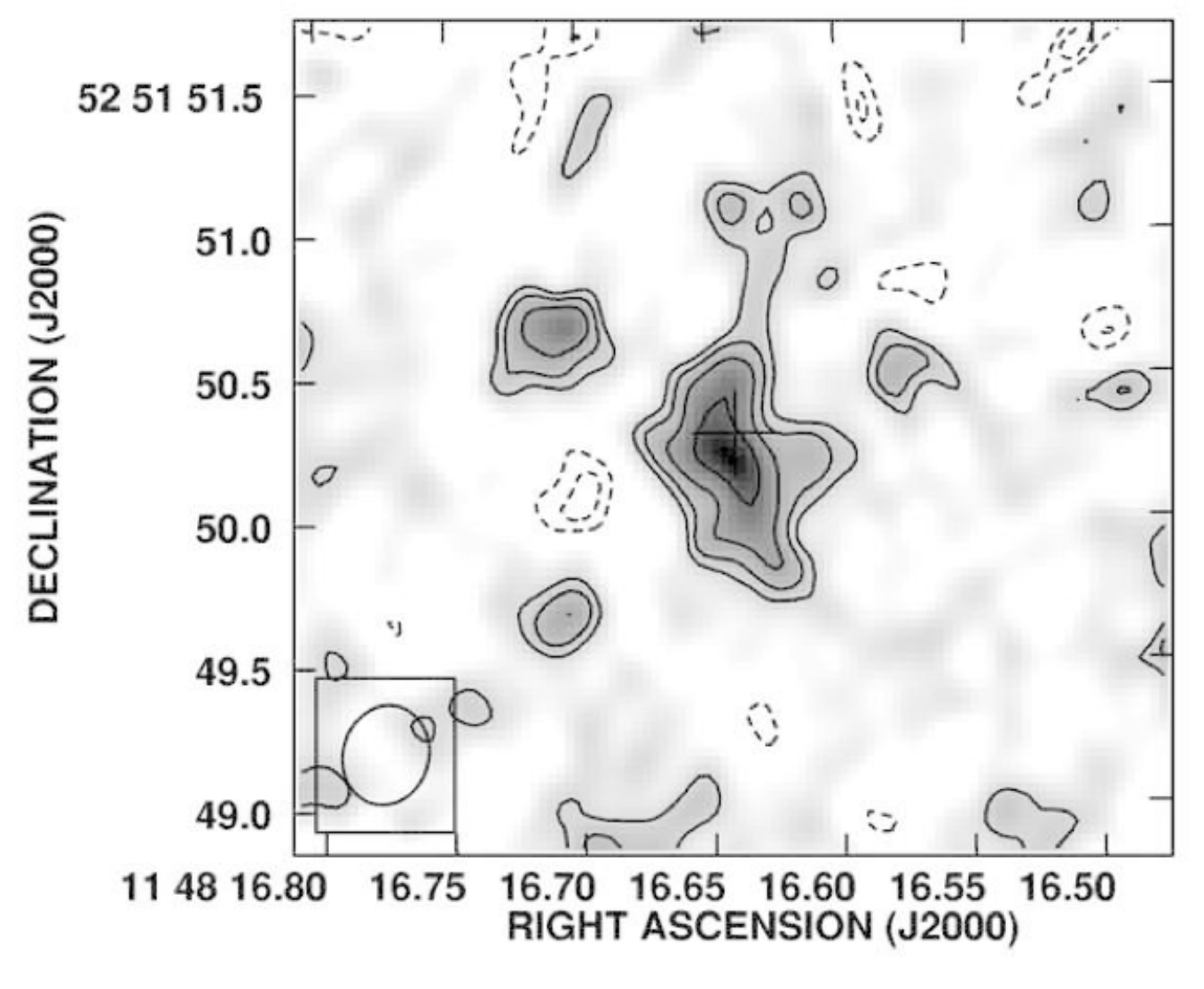}
\caption{VLA redshifted CO $J$=3--2 map of the quasar J1148+5251 using the combined
B- and C-array data sets (covering the 
total bandwidth, 37.5 MHz or 240 km s$^{-1}$), from Walter et al.\ (2004). Contours are 
shown at --2, --1.4, 1.4, 2, 2.8, and 4 $\times \sigma$ (1 $\sigma$ = 43 $\mu$Jy 
beam$^{-1}$). The beam size (0.35\arcsec $\times$0.30\arcsec) is shown in the bottom left corner; the 
plus sign indicates the SDSS position (and positional accuracy) of J1148+5251.}
\label{fig:red1}
\end{figure}

Recently, CO has been detected in galaxies at redshifts $>$6\footnote{Note that interferometers in general have 
an advantage over single-dish telescopes when detecting molecular emission at high redshift since their 
high-resolution imaging capabilities provide the spatial information needed to associate a detection with a 
specific 
object.}.  These and other observations in the cm/mm of $z>6$ galaxies are summarized by  Carilli et al.~(2008; 
see also the large surveys of CO at $z$ $>$ 6 by Wang et al.\ 2010, 2011a and references therein).  Current 
instrumentation sensitivities are such that detections are limited to hyperluminous infrared galaxies, i.e. L$_{\rm 
FIR} > 10^{13}$ L$_\odot$.  Only a small fraction of galaxies are this luminous.  The best-studied such object is 
J1148+5251 with a redshift of $z=6.419$ (see Carilli et al.~2008).  For example, Walter et al.\ (2004) imaged the 
CO $J$=3--2 emission (Figure~\ref{fig:red1}) using the VLA, from which they were able to infer the dynamical 
mass.  Walter et al.~(2009) were not able to detect the [NII] line at  205 $\mu$m,, but did detect the CO $J$=6--5 
transition.  More recently, Wang et al.\ (2011b) detected the lower-energy CO $J$=2--1 transition and Reichers 
et al.\ (2009) imaged CO $J$=7--6 and CI ($^{3}P_{2}$--$^{3}P_{1}$) emission towards this source.  These and 
other (dust continuum) observations show that there was already a significant abundance of metals and dust by 
this epoch.

\begin{figure} 
\includegraphics[scale=0.5]{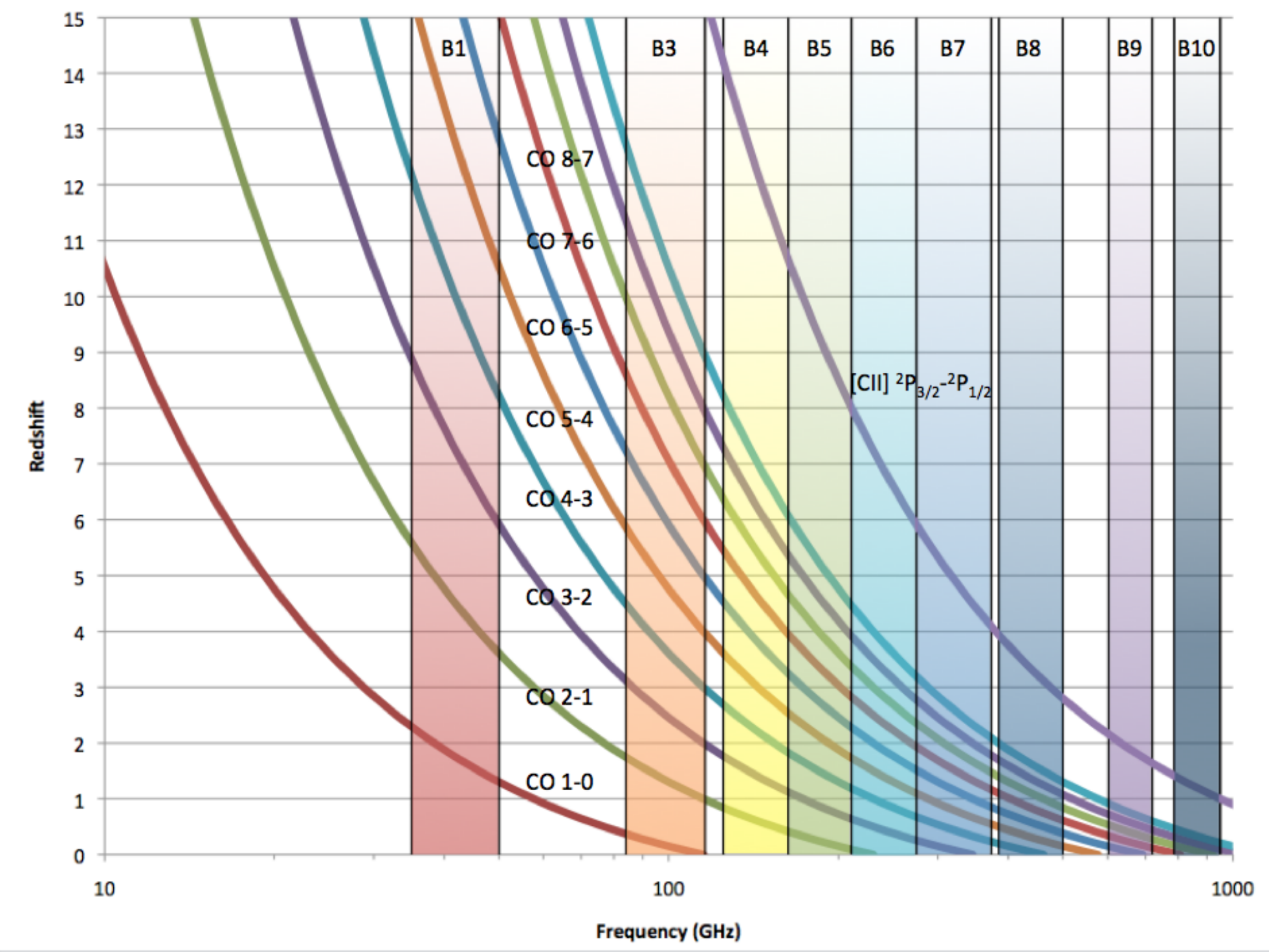}
\caption{Observable frequencies of $^{12}$CO rotational transitions and [CII] $^{2}$P$_{3/2}$--$^{2}$P$_{1/2}$
as a function of redshift.  The frequency ranges of the ALMA Bands are also shown.  Note that the range
for Band 1 reflects the new nominal range of 35-50 GHz.}
\label{fig:red2}
\end{figure}

Figure~\ref{fig:red2} shows the observable frequency of rotational transition sof $^{12}$CO, from $J$=1--0 through $J
$=10--9, as a function of redshift.  Also shown are the frequency ranges of the ALMA Bands (excluding Band 2 for clarity).  
Note that this Figure shows the new nominal range of Band 1 of 35-50 GHz, as this range will yield the highest sensitivities.  
As the Figure shows, Band 1 receivers will be able to detect galaxies in $J$=3--2 at $6 \lesssim z \lesssim 9$, i.e., in the 
redshifts of the era of re-ionization ($z \ga 6$), while higher Bands can only observe higher-$J$ lines that may be less 
excited.  (For example, Band 3 receivers would be able to detect $J$=6--5 emission in the range $4.8 \lesssim z \lesssim 
7.2$.)  Moreover, Band 1 receivers would enable coverage for $J$=2--1 and $J$=1--0 emission at $3.6 \lesssim z \lesssim 
5.6$ and $1.3 \lesssim z \lesssim 2.3$, respectively.  Assuming a 150 $\mu$Jy CO $J$=2--1 line of width
$\sim$600 km~s$^{-1}$ at $z = 5.7$, a 5 $\sigma$ detection would take less than 4 hours with 50 ALMA antennas.
 
Band 1 will also allow multiline observations toward certain subsets of redshifts.  For example, galaxies at $1.3 \lesssim z 
\lesssim 2.3$ can be observed in Band 1 but also at $J$=4--3 and $J$=3--2 in Band 4 (NB: a small gap exists at $z$ $
\approx$ 1.8).  Figure~\ref{fig:red2} also shows that in addition the [CII] $^{2}$P$_{3/2}$--$^{2}$P$_{1/2}$ line can be 
observed toward a subset of these galaxies at $1.6 \lesssim z \lesssim 2.2$ using Band 9.  The [CII] line can also be
observed toward galaxies at $2.8 \lesssim z \lesssim 5.9$ using Bands 7 and 8 (NB: a small gap in redshift coverage
exists at $z$ $\approx$ 4).

As with other ALMA Bands, high-redshift science will be done with Band 1 in a targeted mode, i.e., towards known high-$z$ sources.  An instantaneous $\sim$8 GHz range of frequency coverage, however, will allow
significant sensitivity to other sources proximate on the sky to the known target source but at quite different 
redshifts.  Indeed, ``blank-sky" surveys, made by pointing ALMA towards one location but stepping through the entire Band 1 frequency range, are an enticing possibility (see, e.g., Aravena et al.\ 2012).

\subsubsection{Quasar Host Galaxies}

The discovery of molecular gas in quasar host galaxies at $z \sim 6$, when the Universe
 was less than 1~Gyr old 
(Walter et al.\ 2003; Bertoldi et al.\ 2003; Carilli et al.\ 2007),
has opened a new window on the study of gas in systems that
contributed to the reionization of the Universe. Studies 
of how the molecular gas properties should evolve,
 and how they can be used to reveal the dynamics of these 
 \textit{massive} systems, have recently prompted a new generation of  
 semi-analytic models with the further aim of understanding how 
 high-redshift quasars fit within the context of large-scale structure
formation. 
Li et al.\ (2007, 2008) have used state of the art
 N-body simulations to show that the observed optical properties of high-redshift
 quasars can be explained if these objects formed in the most
 massive dark matter halos ($\sim
 8\times10^{12}$~M$_{\odot}$) early on. These models predict that the 
 most luminous quasars should evolve due to an increase of major mergers, 
which one would expect to find evidence for in the 
CO line profiles and the spatial distribution of the molecular gas
(Narayanan et al.\ 2008). 
Detailed radiative transfer models of the FIR spectral energy distribution of these
systems have been driven by the observations of one 
$z = 6.42$ quasar (namely J1148+5251; Walter et al.\ 2003, 2004).
Larger samples of CO-detected quasars are needed to 
provide better constraints on the models and constrain
dynamical masses to compare with infrared measurements
of black-hole masses (e.g., from MgII lines) and explore the
(possible) evolution of the relation between the masses of 
central black holes and bulges.  Current 3~mm surveys of \hij CO line
emission in $z\sim 6$ FIR-luminous quasars are being conducted
with the PdBI, having successfully detected CO line emission in
eight objects (Wang et al.\ 2010, 2011a). 
  
\subsubsection{Lyman-$\alpha$ Emitters}

The rarity of the
luminous quasars at early times suggests that their UV emission is unlikely to contribute
significantly to the reionization of the Universe (e.g., Fan et al.\
2001). A more important type of galaxy in the context of cosmic re-ionization
are the Lyman-$\alpha$ emitters (hereafter LAEs).  These galaxies were
discovered through their excess emission in narrow-band images centered 
on the redshifted Lyman-$\alpha$ line  (e.g. Hu et al.\ 1998; Rhoads et al.\ 2000;
Taniguchi et al.\ 2005), and constitute a significant fraction of
 the star-forming galaxy population at $z \sim 6$.
 While the star-formation rates in LAEs inferred from their UV
 continuum emission are a few tens of solar masses per year
 (e.g. Taniguchi et al.\ 2005),
 their number density and the shape of the Lyman-$\alpha$ 
emission line provide important probes of physical 
conditions in the Universe around the epoch of reionization. As such,
 it is of fundamental importance that  we understand the properties related to their
 star-formation activity. In particular, we need to quantify the amount of molecular gas available for
 fuel. Wagg, Kanekar \& Carilli (2009) used the Green Bank Telescope to  search for
 \co line emission in $z > 6.5$ LAEs, targeting two sources including the highest
 spectroscopically confirmed redshift LAE at $z = 6.96$ (Iye et al.\
 2006). The limits to the CO line luminosity implied by the
 non-detections of \co in these two objects suggest modest
 molecular gas masses ($\la$\,10$^{10}$~M$_{\odot}$). 
This conclusion, however, is based on observations of only two
 objects, and future studies would benefit from the sensitivity gained
 by observing higher-order CO transitions, whose
 flux density may scale as $\nu^2$ due to a contribution to the
 molecular gas excitation by the cosmic microwave background radiation
 (19~K at $z = 6$).   With other facilities, it has been proven challenging to
 detect even the higher energy $J$=2--1 line from Lyman-$\alpha$-emitting
 galaxies at these redshifts, using existing facilities (Wagg \& Kanekar 2011).
 At these redshifts, such studies would require ALMA, including the Band
 1 receivers.

\section{A Broad Range of Science Cases}
\label{sec:range}

Along with the two science cases presented above, there is a wealth of scientific opportunity available to the ALMA community if the Band 1 Receiver is built. Here we highlight a selection of science cases which would significantly benefit from Band 1 on ALMA.

\subsection{Continuum Observations with ALMA Band 1}

The astrophysical continuum radiation at wavelengths of $\sim 1\,$cm is
relatively unexplored. Yet this radiation is key to understanding radio
emission mechanisms and probing regions that are optically thick at
shorter wavelengths.  The resolution and sensitivity of ALMA Band 1
will allow: (1) a diagnostic of the smallest interstellar dust grains, (2) an 
understanding of the nature of pulsar wind nebulae, (3) the detection 
of radio SNe, with constraints on stellar precursors and remnants, and 
(4) a diagnostic of X-ray binaries.

\subsubsection{Very Small Grains and Spinning Dust} 

\begin{figure}
\epsscale{0.8}
\includegraphics[scale=0.8]{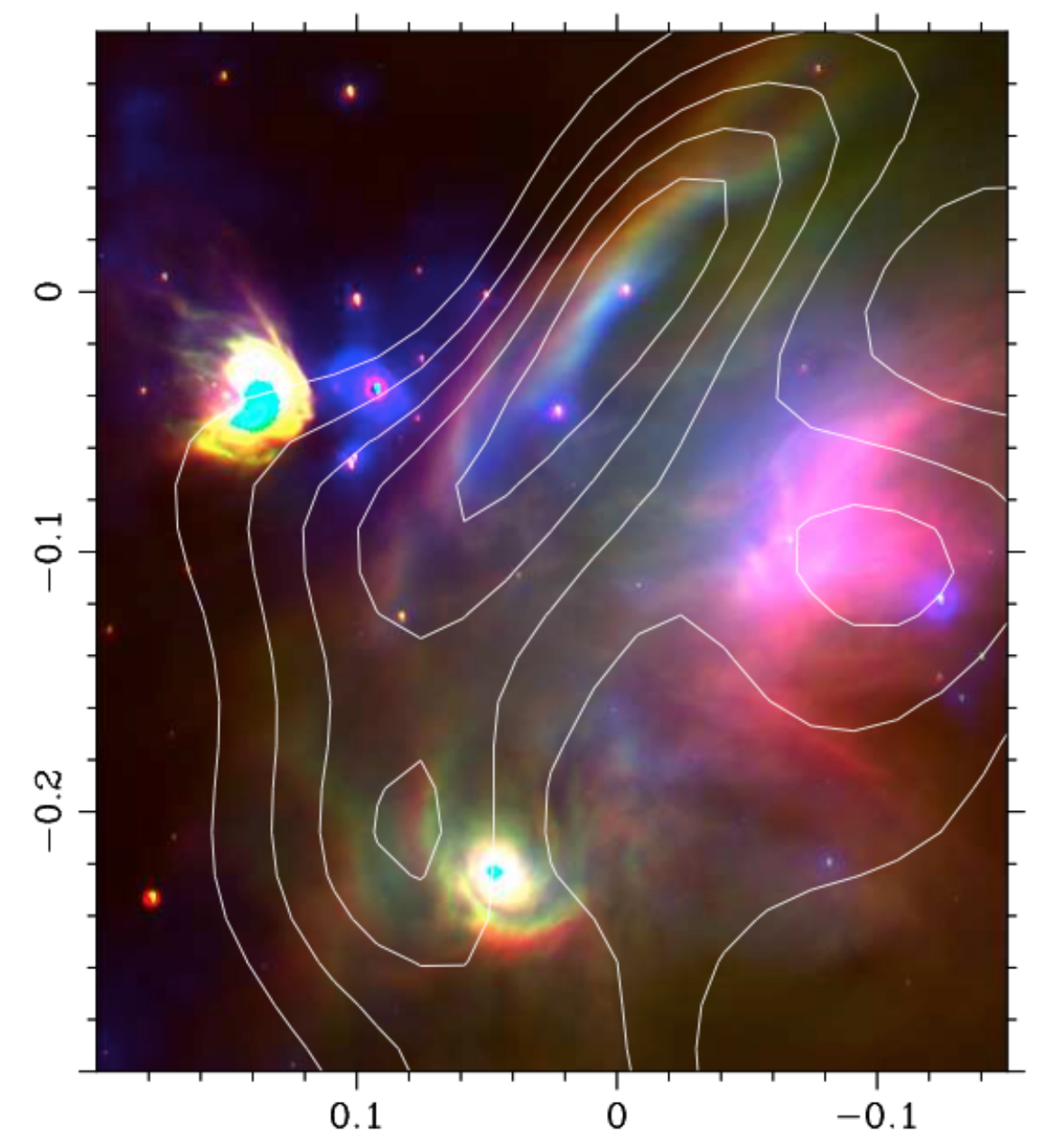}
\caption{Three-colour image of the $\rho$~Oph~W photo-dissociation
  region (Casassus et al.\ 2008). {\bf Red}: MIPS~24~$\mu$m continuum {\bf Green}: IRAC 8~$\mu$m continuum,
  dominated by the 7.7~$\mu$m PAH Band {\bf Blue}: 2MASS~K$_{\rm s}$-band
  image. The $x-$ and $y-$axes show offset in RA and Dec from $\rho$~Oph~W,
  in degrees.  The contours follow the 31~GHz emission, with levels at
  0.067, 0.107, 0.140, 0.170, and 0.197  MJy~sr$^{-1}$.   }
\label{fig:cont1}
\end{figure}

The last decade has seen the discovery of surprisingly bright cm-wavelength 
radio emission in a number of distinct galactic objects but most notably 
in dark clouds (e.g., Finkbeiner et al.\ 2002; Casassus et al.\ 2008 (see 
Figure~\ref{fig:cont1}); Scaife et al.\ 2009a).  The spectrum of this new 
component of continuum radiation can be explained by electric dipole radiation 
from rapidly rotating (``spinning") very small dust grains (VSGs), as 
calculated by Draine \& Lazarian (1998; DL98).  It has also been seen as a 
large-scale foreground in CMB maps, spatially correlated with thermal dust
emission and having a spectrum peaking at $\sim$40 GHz.

All of the existing work aimed at diagnosing this continuum emission is 
derived from CMB experiments on large angular scales, where the bulk of the 
radio signal occurs, e.g., recently by the {\em Planck\ } 
satellite.  Details on small angular scales are crucial, however, for probing
star formation and circumstellar environments.  Simply, progress in the 
understanding of the solid and gaseous states of the ISM requires sufficient 
resolution to separate the distinct environments.   Directly measuring the
VSG abundance and solid state physics is very exciting because VSGs play a
central role in the chemical and thermal balance of the ISM.  For example,
the smallest grains account for most of the surface area available for 
catalysis of molecular formation.

DL98 proposed that the grain size distribution in their spinning dust model 
would be dominated by VSGs, thought to be mostly PAH nanoparticles.  The size 
distribution of VSGs is poorly known as studies of interstellar extinction are 
relatively insensitive to its details.  The existence of VSGs is supported by 
assertions that the existence of a significant amount of carbonaceous 
nanoparticles in the ISM could explain observations of the unidentified IR 
emission features and the strong mid-infrared emission component seen by IRAS 
which must result from starlight reprocessing of ultrasmall grains.  The 
fraction of the ISM carbon content proposed to exist in VSGs considerably 
exceeds that implied by the MRN dust size distribution, which is known 
to underestimate it.  

Observationally determining PAH content in dust clouds is not straightforward.
Where there is a strong source of UV flux present it is possible to identify 
PAHs by their spectral emission features.  In the case of pre-stellar and Class 
0 clouds, however, these features are absent.  With microwave observations from 
ALMA Band 1 constraining the spinning dust SED at similar resolution to, e.g., 
Spitzer or the forthcoming MIRI instrument on the JWST, it will be possible to 
measure the size distribution of VSGs directly from the data.  This work will 
also be important in the context of circumstellar and protoplanetary disks, 
where the proposed population of VSGs may have important implications for disk 
evolution.  Certainly, spinning dust emission will provide a better measure of 
the small grain population within circumstellar disks than PAH emission since 
favorable excitation conditions for PAHs exist only in the outermost layers 
of the disk.  Since all the VSGs in the disk should contribute spinning dust 
emission, such emission will provide a much better probe of the mass in VSGs.  
Combining this information with the PAH emission features would then also give 
us a useful measure of sedimentation in disks.

Theoretically, spinning dust emission from a VSG population has been shown 
to dominate the thermal emission from stellar disks (around HAeBe stars) at significant
factors at frequencies $\leq$ 50 GHz (Rafikov 2006).  The existence of these VSGs
has been confirmed observationally using PAH spectral features as seen in the 
disks of Herbig Ae/Be stars (Acke \& van den Ancker 2004) but it has not been 
detected in protoplanetary disks due to a lack of strong UV flux.  Since 
spinning dust emission has been observed to be spatially correlated with PAH 
emission (Scaife et al.\ 2009b) spinning dust may provide a unique window on 
the small grain population of these disks.  In the context of disk evolution, these
recent measurements conflict with the established 
view that dust grains are expected to grow as disks age.  It may be the case 
that dust fragmentation is important in disks (Dullemond \& Dominik 2005), or 
there exists a separate population of very small carbonaceous grains distinct from 
the MRN distribution (Leger \& Puget 1984; Draine \& Anderson 1985).  This 
second proposition has not only important implications for the study of 
circumstellar disks but also for the complete characterization of dust and 
the ISM more generally.  

The arcsecond resolution necessary for these measurements will be achievable 
with several ALMA configurations and Band 1.  From the models of Rafikov 
(2006), the difference between a thermal dust spectrum with $\beta$ $\approx$ 
1 and the predicted spinning dust contribution for a brown dwarf disk would be 
observable at 5 $\sigma$ in a matter of minutes with ALMA Band 1.  With longer 
observation times and consequently higher sensitivity, it will also be possible 
to distinguish between different grain size distributions and physical 
conditions within the disk (such as grain electric dipole moments, rotational 
kinematics, optical properties and catalysis of molecule formation).

Spinning dust emission provides a unique insight into the VSG population under 
conditions where it is not possible to observe using mid-IR emission.  The high 
resolution and excellent sensitivity of ALMA are ideal for differentiating the 
distinct environments where the VSG population resides and will be crucial for 
probing star formation and circumstellar regions.  Band 1 will allow routine 
surveys of the new continuum component at its spectral maximum.

\subsubsection{Pulsar Wind Nebulae}  

Pulsars generate magnetized particle winds that inflate an expanding 
bubble called a pulsar wind nebula (PWN) whose outer edge is confined by 
the slowly expanding supernova ejecta. Electrons and positrons are 
accelerated at the termination shock some $0.1\,$pc distant from the 
pulsar. Those relativistic particles interact with the magnetic field inside the 
wind blown bubble to produce synchrotron emission across the entire 
electromagnetic spectrum. Particles accelerated at the shock form 
toroidal structures, known as wisps, and some of them are collimated 
along the rotation axis of the pulsar, contributing to the formation of 
jet-like features. The synchrotron emission structure in the post-shock 
and jet regions provide direct insight on the particle acceleration 
process, magnetic collimation, and the magnetization properties of the 
winds in PWNe. These observations have so far (except for the Crab 
Nebula) been limited to the X-ray band with the {\em Chandra} satellite
(e.g., Helfand et al.\ 2001).

\begin{figure}
\includegraphics[scale=0.8]{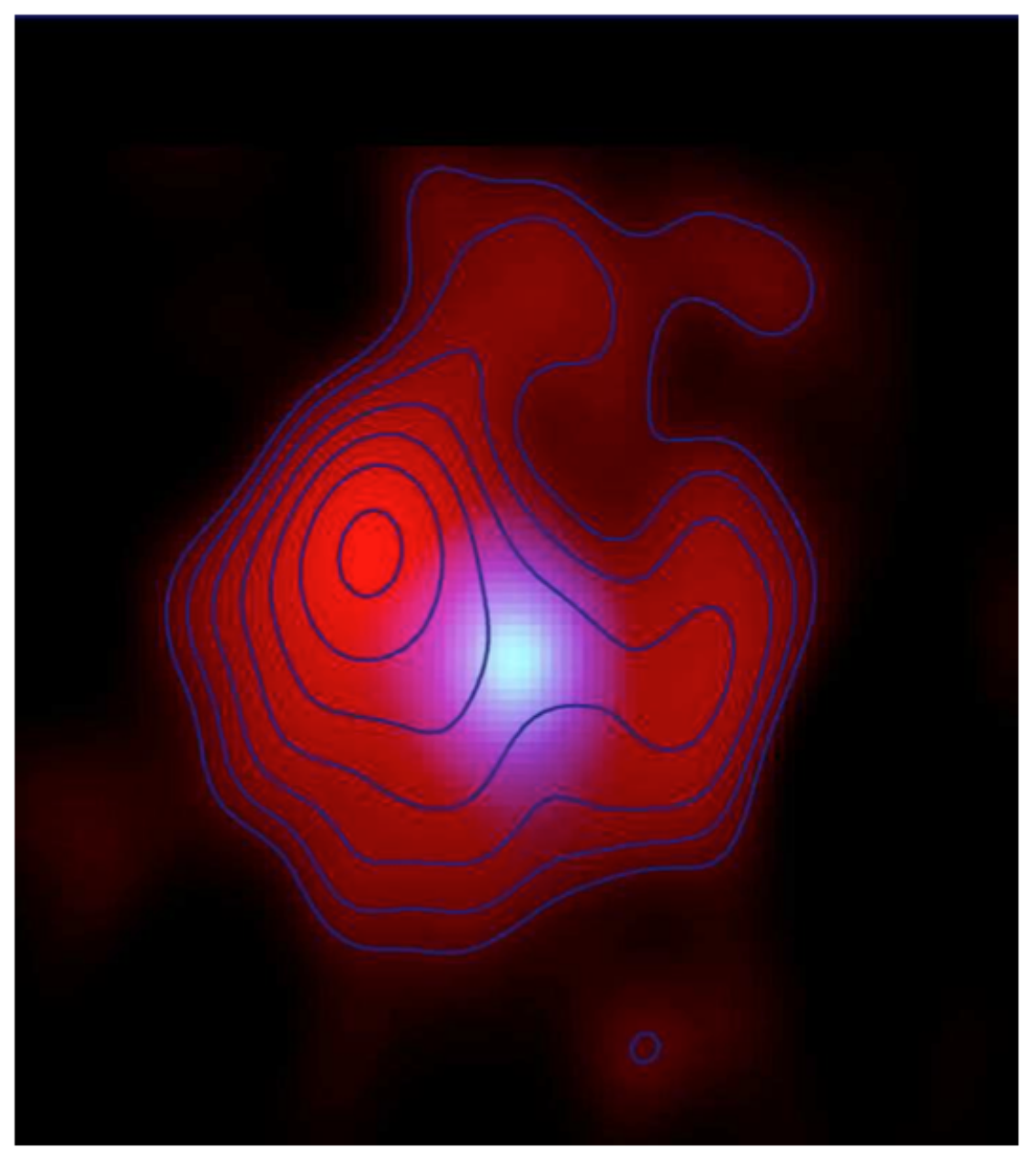}
\caption{
Two-colour VLBI image of SN 1986J highlighting the emergence of a central
component. The red colour and the contours represent the 5.0 GHz radio
brightness. The contours are drawn at 11.3, 16. 22.6É90.5\% of the peak
brightness of 0.55 mJy/bm. The blue to white colours show the 15 GHz brightness
of the compact, central component. The scale is given by the width of the
picture of 9 mas. North is up and east to the left. For more information on the emergence 
of the compact source, see Bietenholz et al.\ (2004). }
\label{fig:cont2}
\end{figure}

ALMA 
has the sensitivity and resolution necessary to detect PWNe features 
at high radio frequencies, where we can detect the emission from 
relativistic particles that have much longer lifetimes than in
X-rays.
At cm/mm-wavelengths, flat-spectrum synchrotron PWNe
stand out over steep-spectrum SNRs 
[e.g.\ as seen in the Vela PWN, Hales et al.\ (2004) and also as discussed
in Sec.~4.1.3 below, and illustrated in Fig.~\ref{fig:cont2}, taken from Bietenholz
et al.\ (2004)]
with minimal
confusion from the Rayleigh-Jeans tail of submm dust. ALMA Band 1
will allow observations in the frequency regime where PWNe dominate, and
bridge an important gap in frequency coverage, where spectral features
such as power-law breaks occur and linear polarization observations
do not suffer from significant Faraday rotation.

\subsubsection{Radio Supernovae} 

Radio supernovae occur when the blast wave of a core-collapse supernova (SN)
sweeps through the slowly expanding wind left over from the progenitor red
supergiant. Particle acceleration and magnetic field amplification lead to
synchrotron radiation in a shell bounded by the forward and reverse shocks
(Chevalier 1982).  In general, free-free absorption of the radiation in the
ionized foreground medium coupled with the expansion of the SN causes the radio
light curve first to rise at high frequencies and subsequently at progressively
lower frequencies while the optical depth decreases. When the optical depth has
reached approximately unity, the radio light curve peaks and decreases
thereafter (e.g., Weiler et al.\ 2002). These characteristics allow estimates to
be made of the density profiles of the expanding ejecta and the circumstellar
medium and also of the mass loss of the progenitor. Resolved images of SNe
provide information, e.g., on the structure of the shell, size, expansion
velocity, age, deceleration, and magnetic field, in addition to refined
estimates of the density profiles and the mass loss (Bartel et al.\ 2002).
Radio observations of SNe can be regarded as a time-machine where the history of
the mass loss of the progenitor is recorded tens of thousands of years before
the star died.
Finally, the SN images can be used to make a movie of the expanding shell of
radio emission and to obtain a geometric estimate of the distance to the host
galaxy (Bartel et al.\ 2007).

ALMA Band 1 will allow exciting science to be done in the areas of radio light
curve measurements, imaging of a nearby SN and, in conjunction with VLBI,
imaging of more distant SNe. Depending on the medium, the delay between the
peak of the radio light curve at 20 cm and 1 cm can be as long as 10 years, as
for instance was the case of SN 1996cr (Bauer et al.\ 2008). Absorption
can also occur in the source itself. In case of SN 1986J, a new component
appeared in the radio spectrum and in the VLBI images about 20 years after the
explosion and then only at or around 1 cm wavelength. The component is located
in the projected center of the shell-like structure of the SN and may be
emission from a very dense clump fortuitously close to that center, or possibly
from a pulsar wind nebula in the physical center of the shell (Figure~\ref{fig:cont2},
Bietenholz et al.\  2004, 2010).  Observations in Band 1 minimize the absorption effect
relative to observations at longer wavelengths and thus allow investigations of
SNe at the earliest times without compromising too much on the signal to noise
ratio of a source with a steep spectrum. ALMA with Band 1 has the sensitivity
to measure the radio light curves of 10s to 100 SNe. In addition, ALMA with
Band 1 may be particularly sensitive in finding ÒSN factoriesÓ in starburst
galaxies (e.g., Lonsdale et al.\ 2006) where relatively large opacities would
otherwise hinder or prevent discovery.

ALMA with Band 1 will allow high-dynamic range images of SN 1987A in the Large
Magellanic Cloud with a resolution of about 300 FWHM beams across the area of
the shell in 2014.  Such data would be a significant improvement over presently
obtainable images (Gaensler et al.\ 2007; Laki{\'c}evi{\'c} et al.\ 2012). Also, since the size of the SN increases
by one Band 1 FWHM beam width per 3 years, the expansion of the shell can be
monitored accurately and in detail, making this SN an important target for ALMA.

ALMA with Band 1 could also be an important element of VLBI arrays. 
SN VLBI observations at
1 cm wavelength have provided clues about physical conditions at the earliest
times after the transition from opaqueness to transparency, and SN VLBI with
Band 1 will focus on this area of research.

\subsubsection{X-ray Binaries}

X-ray binaries (i.e., binary star systems with either a neutron star or a black 
hole accreting from a close companion) frequently show jet emission.  Most of 
these systems are transients. Typically, 1-2 black hole X-ray binaries undergo 
a transient outburst per year, while neutron stars outburst at a slightly higher 
rate.  Outbursts typically last several months (although there are some which
are both considerably longer or shorter), and during outbursts, X-ray luminosities
can change by as much as 7 orders of magnitude. The radio luminosities of
systems seen to date correlate well with the hard X-ray luminosities (i.e., the
luminosity above $\sim$20 keV), albeit with considerable, yet poorly understood
scatter.

When the X-ray spectra become dominated by thermal X-ray emission, the radio 
emission often turns off (e.g., Tananbaum et al.\ 1972; Fender et al.\ 1999), 
but the extent to which the flux turns down is still poorly constrained. This 
turndown is not seen in neutron star X-ray binaries (Migliari et al.\ 2004). 
The reduced radio emission in black hole X-ray binaries when they have soft 
X-ray spectra can be explained by models of jet production in which the jet 
power scales with the polodial component of the magnetic field of the accretion 
flow (e.g., Livio, Ogilvie \& Pringle 1999), and may have implications for 
the radio loud/quiet quasar dichotomy (e.g., Meier 1999; Maccarone, 
Gallo \& Fender 2003).  The still-present radio emission from neutron stars 
in their soft state may be indicating that the neutron star boundary layers 
play an important role in powering jets (Maccarone 2008).  The soft states
of X-ray transients are short-lived.  During them, there may be decaying 
emission from transient radio flares launched during the state transitions. 
Therefore, to place better upper limits on the radio jets produced during 
the soft state, a high sensitivity, high frequency system with a very high duty
cycle is needed.

The radio properties of X-ray binaries with neutron star primaries are much 
more poorly understood than those of black hole X-ray binaries.  This situation
is partially because the neutron star X-ray binaries are fainter in X-rays than 
are the black hole X-ray binaries. There is, however, additionally some 
evidence that neutron star X-ray binaries show a steeper relation between X-ray 
luminosity and radio luminosity than do the black hole X-ray binaries, with 
$L_R \propto L_X^{0.7}$ for the black holes and $L_R \propto L_X^{1.4}$ for 
the neutron stars.  This difference may be explained if the neutron stars are 
radiatively efficient (i.e., with the X-ray luminosity scaling with the accretion
rate) while the black holes are not (i.e., with the X-ray luminosity scaling with the
square of the accretion rate, as has been proposed by Narayan \& Yi 1994) -- 
see Koerding et al.\ (2006). Radio/X-ray correlations for neutron star X-ray
binaries are, to date, based on small numbers of data points from few sources,
and the most recent work (Tudose et al.\ 2009) indicates that the situation may
be far more complex than the picture presented above. 
 

\subsubsection{Spatial and Flaring Studies of Sgr A*}

Near-IR and radio observations provide compelling evidence that the compact nonthermal radio source Sgr A* is identified with 
a 4 $\times$ $ 10^6$ solar mass black hole at the center of the Galaxy (Reid and Brunthaler 2004; Ghez et al. 2008; Gillessen 
et al. 2009).  It is puzzling, however, that the bolometric luminosity of Sgr A* due to synchrotron thermal emission from hot 
electrons in the magnetized accretion flow is several orders of magnitude lower than expected from the accretion of stellar 
winds. There have been two different approaches to address this puzzling issue.  One is to search for the base of a jet from 
Sgr A* and identify interaction sites of a jet with the ionized and molecular material surrounding Sgr A*.  The other is 
to study the correlations of the variable emission from Sgr A* at centimeter and millimeter bands.  Studies of images and variability are well suited using ALMA's Band 1 and will be complementary to each other in addressing the key question as to why Sgr A* is so 
underluminous.

\subsubsubsection{A Jet from Sgr A*} 

\begin{figure}
\begin{center}
\includegraphics[scale=0.65]{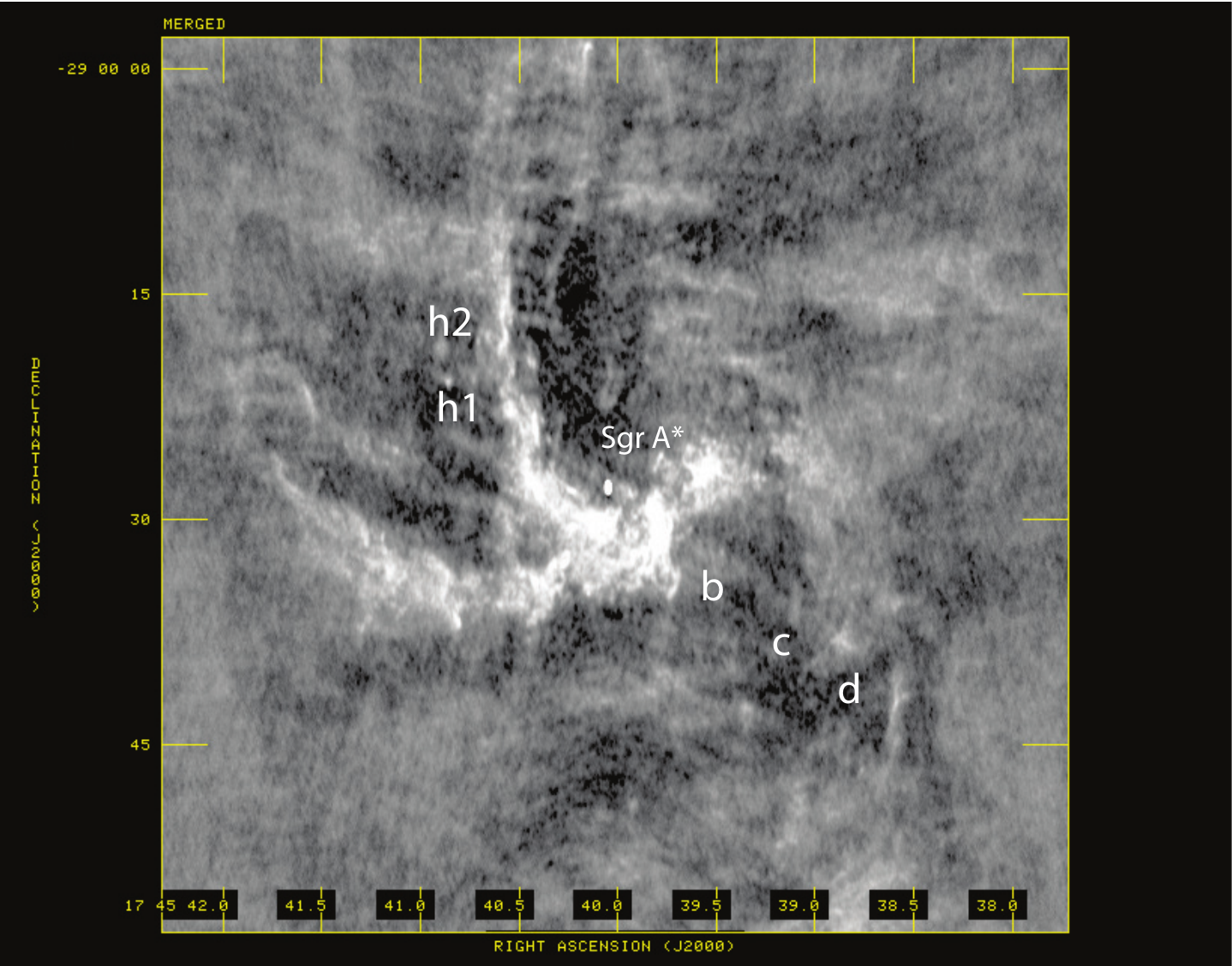}
\caption{ {\it (a) Left} A 22 GHz image of $0.36''\times0.18''$ resolution (PA=2$^\circ$)  constructed by
 combining JVLA A- and B- array data.}
\label{fig:sgra22}
\end{center}
\end{figure}

Recent JVLA observations at radio wavelengths presented a tantalizing detection of a jet-like linear feature appearing to emanate from Sgr A* (Yusef-Zadeh et al. 2012).  Figure~\ref{fig:sgra22} shows a grayscale 23 GHz image of the inner 30$''$ of Sgr A*. A 
new linear feature is noted running diagonally crossing the bright N and W arms of the mini-spiral, along which several 
blobs (b, c, d, h1 and h2) are detected. What is interesting about the direction in which the linear feature is detected is 
that several radio blobs have X-ray and FeII/III counterparts also along the axis of the linear structure. In 
addition, the extension of the linear feature appears to be polarized at 8 GHz, suggesting that this feature is a synchrotron 
source. The radio-polarized linear jet-like structure is best characterized by a mildly relativistic 
jet-driven outflow from Sgr A*, and an outflow rate $\gamma\dot{M}\sim 10^{-6}$ \msol\, yr$^{-1}$.

The linear arrangements of antennas in the JVLA configurations can lead to linear structures in the residual beam pattern due to deconvolution errors.  ALMA's configurations, however, should lead to data with better, more-uniform uv coverage and will establish the reality of the linear structure.  In particular, Band 1 will be most effective in studying the
faint jet-like feature from Sgr A*.  Dust emission from the immediate environment of Sgr A* dominates fluxes at shorter wavelengths relative to optically thin non-thermal emission from the jet with a steep energy spectrum.  
Thus,  observations with Band 1 are critical for  measuring properly the morphology, spectral index and polarization characteristics of the jet emanating from Sgr A*.  Although, Sgr A* is a unique object in the
Galaxy, similar motivations also apply to other non-thermal radio continuum sources such as microquasars,
e.g., 1E1740.7-2942, having faint radio jets and are located in the inner Galaxy. 

\subsubsubsection{Time Delay at Centimeter Wavelengths}
 
\begin{figure}
\begin{center}
\includegraphics[scale=0.5]{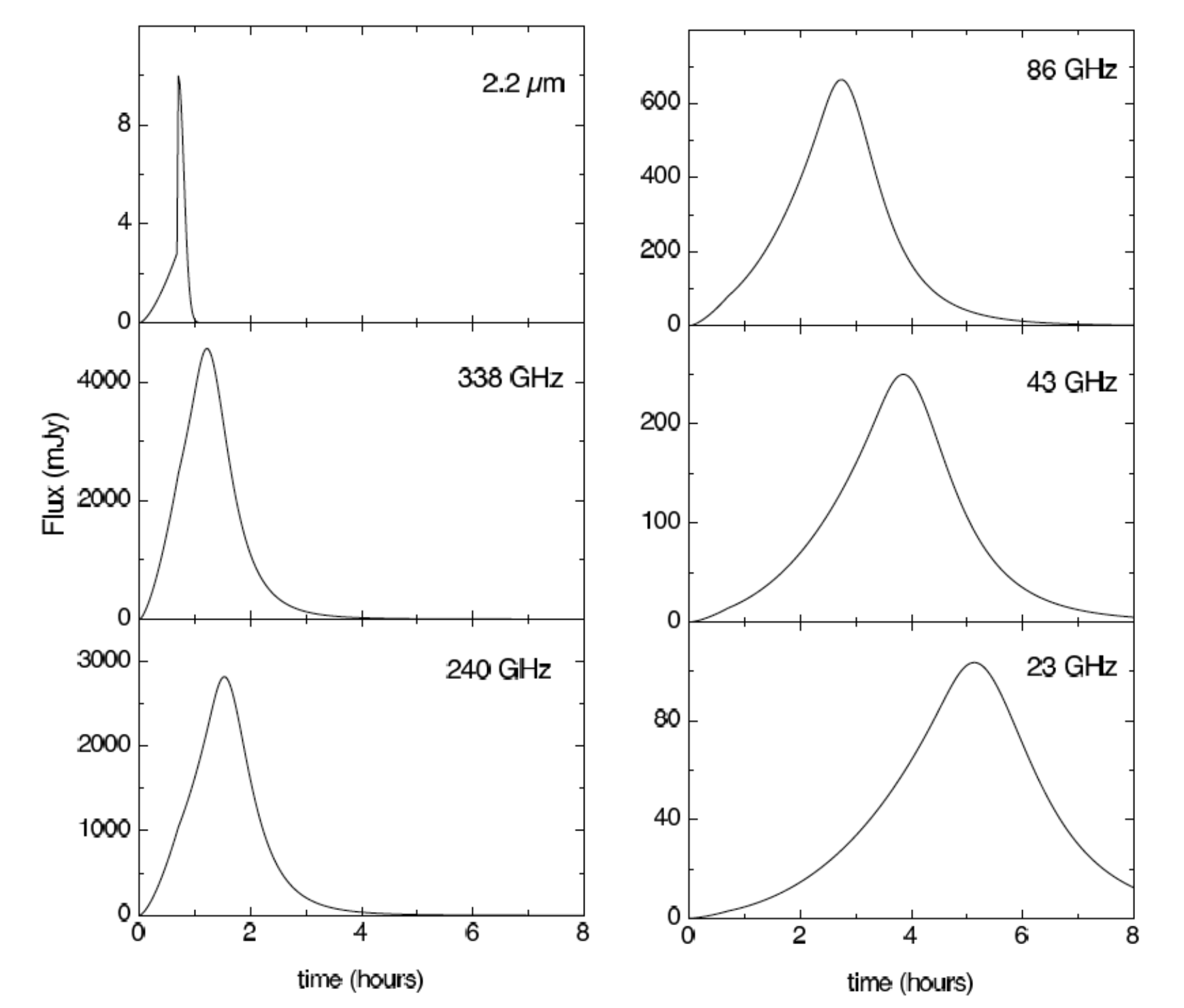}
\caption{ {\it} 
Theoretical light curves of Stokes I for
  optically thick synchrotron emission at five different bands
  corresponding ALMA Bands 3, 6, 7 and 9 as a function of expanding
  blob radius.  These light curves assume an energy power law index
  p=1 where n(E)$\propto$ E$^{-p}$.  
 }
\label{fig:synchro}
\end{center}
\end{figure}

Recent radio measurements have detected
a time delay of $\sim$30 $\pm$ 10 minutes between the peaks of 7 mm and 13 mm radio continuum emission
 toward Sgr A* (Yusef-Zadeh et al.\ 2006).  This behavior is consistent with a picture of a flare in which the synchrotron emission is initially optically thick.  Flaring at a given frequency is produced through the adiabatic expansion of an initiallyoptically thick blob of synchrotron-emitting relativistic electrons.  The intensity grows as the  blob expands, then peaks and declines at each frequency that the blob becomes optically thin. 
This peak first occurs at 43 GHz and then at 22 GHz about 30 minutes later.   Theoretical light curves of flare emission, as shown in Figure~\ref{fig:synchro}, show that flare emission occurs at high near-infrared frequencies first and is increasingly delayed at successively lower ALMA frequencies that are initially optically thick. 

The limited time coverage of JVLA observations at radio wavelengths means that there can be a large uncertainty in determining the underlying background flux level of a particular flare, as well as difficulty identifying flares in different bands. Observations of Sgr A* with a long time coverage using ALMA's Band 1 
can fit the corresponding light curves simultaneously to place much tighter constraints on the derived physical parameters of the flare emission region. Two of the parameters that are of interest are the expansion speed of the hot plasma and the initial magnetic field. These quantities characterize the nature of outflow and cooling processes relevant to millimeter emission.  The fitting of a light curve at one frequency will automatically generate models for any other frequency.  We should be able to test the time delay between the peaks of flare
emission within Band 1. 

What has emerged from past observing campaigns to study Sgr A* is that radio, submillimeter, near-infrared, and X-ray emission can be powerful probes of the evolution of the emitting region since they are all variable.
We now know that flare emission at infrared wavelengths is due to optically thin synchrotron emission that is detected when a flare is launched 
(Eckart et al. 2006).  The
relationship between radio and near-infrared/X-ray flare emission has remained unexplored due the very limited simultaneous time
coverage between radio and  infrared telescopes. The continuous variations of the radio flux on hourly time scale also
makes the identification of radio counterparts to infrared flares difficult. In spite of the limited 
coverage in time, the
strong flaring in near-infrared/X-ray wavelengths has given us an opportunity to examine if there is a correlation with variability
at radio frequencies.  A key motivation for observing Sgr A* is to compare its flaring activity with the adiabatic expansion picture. One of the prediction of this model is a time delay between the peaks of optically thin 
near-infrared emission and optically thick radio emission, as discussed above. 
From this model,  a near-infrared flare of short duration of 0.5-1 hr is expected to have a radio
counterpart shifted in time by 3-5 hr of duration of $\sim2$ hr.   

\begin{figure}
\begin{center}
\includegraphics[scale=0.6,angle=0]{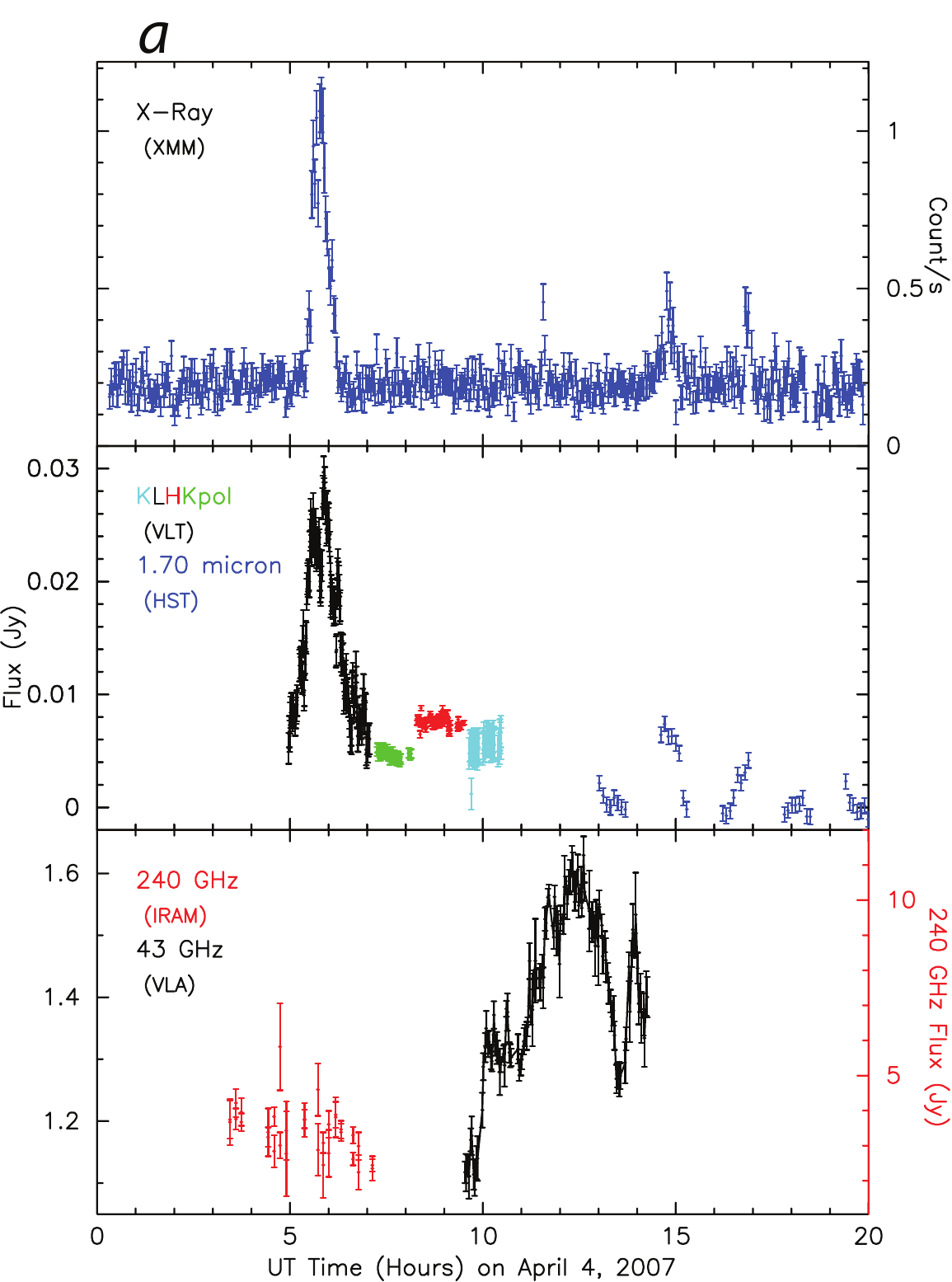}
\caption{The light curves of
Sgr A* on 2007 April 4 obtained with  XMM in X-rays (top), VLT and HST in NIR
(middle), and IRAM-30m  and  VLA at 240 GHz and 43 GHz, respectively (bottom).
The NIR light curves in the middle panel are represented as H (1.66 $\mu$m) in red,
K$_s$ and K$_s$-polarization mode
(2.12 $\mu$m) in green and light blue, respectively,  L' (3.8$\mu$m) in black
(Dodds-Eden et al. 2009), and
NICMOS of HST in blue at 1.70 $\mu$m. In the bottom panel, red and black colors
represent the 240 GHz and 43 GHz  light curves, respectively.
 }
\label{fig:sgralc}
\end{center}
\end{figure}

Figure~\ref{fig:sgralc} shows composite light curves of Sgr A* obtained with XMM, VLT, HST, the IRAM 30-m Telescope, and the VLA on 2007 April 4.  These curves reveal that there was no significant variation at 240 GHz during the period when the strong near-infraredX-ray flare took place.  The IRAM observation shows an average flux of 3.42 Jy $\pm$ 0.26 Jy between 5 hr and 6h UT when the powerful near-infrared flare took
place. The millimeter flux is mainly arising from the quiescent component of Sgr A*.  Comparing the light curves of the 43 GHz and 240 GHz data, there is no evidence for a simultaneous radio counterpart to the near-infrared/X-ray flare with no time delays.  Given the limited coverage in time with the VLA, it is clear that we can not be confident about the time delay between radio and near-infrared/X-ray peaks.  There is also no overlap in time between the VLA and Subaru data to test the adiabatic picture of flare emission by making simultaneous NIR and radio observations.  In future, ALMA and VLT will have the best time overlap to test this important aspect of flare emission from Sgr A*.  Although Sgr A* is a unique object in the Galaxy, similar arguments could be 
made for numerous transient sources found in the inner Galaxy. 
 
\subsubsubsection{Close Encounters of Gas Clouds with Sgr A*}

A 3 M$_{Earth}$ cloud of ionized gas and dust named G2 has been recently determined to be on a collision 
course with Sgr~A*.  VLT observations indicate that the G2 cloud approaches pericenter in mid-2013 and it 
will be disrupted and portions will likely be accreted by the massive black hole residing there (Gillessen et al. 
2012).   At the pericenter distance, the velocity of the gas cloud will be 5400~km~s$^{-1}$.  Accordingly, the 
cloud is expected to produce a bow shock that can easily accelerate electrons into a power-law distribution of index $p = 2.5-3.5$, assuming standard shock conditions (Narayan et al. 2012).  Depending on $p$, the 
expected additional emission from Sgr A* ranges from 0.6~Jy to 4~Jy, over a dynamical timescale of $\sim$6
months.  The model behind the additional radio emission from the disruption of G2 by the black hole could have been tested directly with ALMA Band 1 observations.  Though Band 1 receivers will not be ready for the interaction of G2 with Sgr A* by 2013, this close encounter is likely not an isolated event, and future disruptions  of other, similar clouds in the Sgr A* region by the black hole could be monitored with Band 1.

\subsubsection{The Sunyaev-Zel'dovich Effect}

Much of what we know about galaxy clusters has come from X-ray
observations of thermal bremsstrahlung emission of the intra-cluster
medium (ICM).  For example, the angular resolution of {\it Chandra}
has been crucial to advancing our understanding in this area and has
resulted in a renaissance in astrophysical studies of galaxy
clusters.  In recent years, the Sunyaev-Zel'dovich Effect (SZE) has
provided an increasingly important view of these cosmic structures
(Birkinshaw 1999).  Since the SZE signal is proportional to the
product of the electron density and its temperature ($\sim n_e \,T_e$,
compared to $n_e^2\sqrt{T_e}$ for the x-rays), it gives a complementary
view of the physical state of the ICM, one more sensitive to hot phases
that also directly measures local departures from thermal pressure
equilibrium.  To date, the majority of SZE observations have been
carried out at comparatively low angular resolution (beams $> 1'$ in
size), yielding information about the overall bulk cluster properties.
Advances in instrumentation have begun making higher angular
resolution measurements of the SZE possible, revealing previously
unsuspected shock-heated gas in the ICM of clusters previously
thought to be dynamically relaxed (Komatsu et al.\ 2001, Kitayama
et al.\ 2004, Mason et al.\ 2010, Korngut et
al.\ 2011, Plagge et al.\ 2012).  These $10''$ to $20''$ SZE images are
the current state of the art.  A Band 1 receiver suite on ALMA will
surpass this benchmark, making possible detailed studies of the ICM
using the SZE on larger samples and with greater sensitivity than before.

ALMA Band 1 will be capable of addressing a wide range of basic
questions about the observed structure and evolution of clusters.
For example, what is the structure of ICM shocks and the mechanism(s)
responsible for converting gravitational potential energy into thermal
energy in the ICM (Markevitch et al.\ 2007, Sarazin et al.\ 1988)?  What
is the influence of Helium ion sedimentation within the cluster atmosphere
(Ettori et al.\ 2006)? What is the nature of the AGN-inflated ``bubbles'' seen
in the cores of some clusters (Pfrommer et al.\ 2005), and what is the role
of cosmic rays in the ICM? What is the nature of the underlying ICM
turbulence (e.g., Kolmogorov versus Kraichnan)? A particularly rich area
will be the detailed study of ICM shocks, which are common since infalling
sub-clusters are typically transsonic.  Several galaxy cluster mergers have
been observed recently with Chandra and XMM in X-rays with resolutions at
the arcsecond level where substructures become visible (Markevitch, et
al.\ 2000, 2002).  The features of interest for these studies will typically fit
within one or a few ALMA Band 1 fields-of-view and require longer integrations
(several to $\sim$10 hours per pointing).  Note that Band~1 also may have
the sensitivity to detect the SZE from the halos of massive individual ellipticals
or massive groups.

Another important area where high-resolution SZE imaging will have an
impact is the interpretation of SZE survey data. ACT (Dunkley et al.\ 2011),
SPT (Williamson et al.\ 2011), and PLANCK (Planck Collaboration, 2011)
have all conducted $1000+ \, {\rm deg^2}$ surveys to detect and
catalog galaxy clusters via the SZE.  These surveys provide unique and
valuable information about cosmology but their interpretation depends
upon assumptions about the relationship between the SZE signal and
the total virial mass of the halos observed. It is known that both gravitational
(cluster merger) and non-gravitational processes (AGN and supernova
feedback, bulk flows\footnote{By bulk flow, we refer to the motion of a 
cluster itself through its surrounding medium, producing a kinematic
contribution to the observed SZE signal; in theory, this contribution has
a different spectral dependence than the thermal SZE and may be 
distinguishable with good spatial coverage.}, cosmic ray pressure)
give rise to considerable scatter and potential biases (e.g., Morandi
et al.\ 2007) in this relationship. 
Cluster mergers have a particularly dramatic effect on
the SZE, typically generating transsonic (Mach $ \sim 2$-$4$) shock
fronts which can enhance the peak SZE in the cluster by an order of
magnitude (Poole et al.\ 2007, Wik et al.\ 2008).

The systematic
astrophysical uncertainties just described are the limiting factor in making
cosmological inferences from the small published samples of a few
dozen SZE-selected clusters (e.g., Sehgal et al.\ 2011).  ALMA Band 1
is the only foreseen prospect for efficient high-resolution observations
of the large southern hemisphere samples of SZE-selected cluster that
will directly improve inferences from these surveys.  It will image (at
$5''-10''$ resolution) galaxy clusters discovered in the
low-resolution ($\sim 1'$) surveys, detecting shocks and mergers and
identifying ICM substructure, and providing a direct, phenomenological
handle on important survey systematics.  Indeed,
the sensitivity and resolution of an ALMA Band 1 receiver suite allows for
efficient follow-up observations of cluster detections made by blind southern
hemisphere SZE surveys. Thus a study of the cluster selection of these survey
experiments in a statistical manner becomes feasible and new important
insights into the mass-observable relation and its scatter and dependence
on cluster physics can potentially be obtained. The ability to understand
cluster selection in detail is essential to derive reliable constraints on
cosmological models from SZE cluster surveys (see e.g., Geisbuesch
et al.\ 2005; Geisbuesch \& Hobson 2007).

The coming decade will also see an explosion of optical and X-ray
cluster data.  The German/Russian satellite {\it eRosita}, due
to launch in 2014, will carry out the first all-sky X-ray survey since ROSAT
(Merloni et al.\ 2012).  Among other things, it is expected to catalog
$\sim$100,000 clusters out to $z=1.3$ (Cappellutti et al.\ 2011).  Also, the
Dark Energy Survey (DES; Dark Energy Survey Collaboration 2005) is a
$5,000 \, {\rm deg^2}$, mostly southern sky survey also expected to find
$\sim$100,000 galaxy clusters. Targeted SZE observations with Band 1
will be invaluable to determine the properties of clusters at redshifts where
X-ray spectrscopy and gravitational lensing begin to fail.  These
high-$z$ clusters, such as the ACT-discovered SZE cluster ``El Gordo''
at $z=0.89$, weighing in at $M=(2.16 \pm 0.32) \times 10^{15}$ M$_{\odot}$
(Menanteau et al.\ 2011), offer leverage on so-called ``pink elephant''
tests capable of constraining cosmological or gravitational theories
based on the existence of individual extreme objects, i.e., provided their
properties are accurately determined.  It is worthy of note that in
addition to the high-resolution capability, a Band 1 equipped ALMA
Compact Array (ACA) will be comparable in capability to the OVRO/BIMA
arrays which have been used in the current decade to measure the bulk
SZE properties of large northern hemisphere cluster samples (Bonamente
et al.\ 2008). Extending this capability to the southern hemisphere over
the next decade is important to realize the full potential of these
rich cluster samples.

Given the large number of ALMA baselines, the resulting high image
fidelity and dynamic range of the data will be advantageous to SZE studies,
in particular the detailed ones.  In addition, long baseline data from ALMA
can be used to  remove accurately intrinsic and background (i.e.,
gravitationally lensed) discrete source populations. These latter objects
are a signal of substantial interest from another point of view, but they
also set a significant ``confusion noise'' floor to millimeter single-dish
observations, especially considering the factor of $2-3$ boost in source
confusion in clusters due to gravitational lensing (Blain et al.\ 2002).

\begin{figure}
\begin{center}
\includegraphics[scale=0.65]{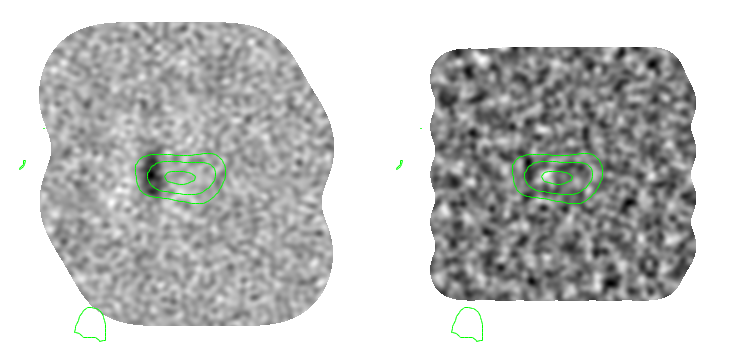}
\caption{Simulated $1.5$~hour ALMA Band 1 (left) and Band 3 (right)
  observations of a galaxy cluster covering $5' \times 5'$. The shock
  is represented as a Gaussian component $5'' \times 25''$ in extent
  with a peak SZE of $y=10^{-4}$, considerably weaker than the
  amplitude observed in RXJ1347-1145 by Mason et al.\ (2010). The Band
  3 data were tapered to the innate resolution of the Band 1 map,
  $\sim 10''$ (FWHM). ACA baselines were not included in this
  simulation but the overplotted contours show the ACA Band 1 image
  (using a $45''$ taper) of the bulk ICM in this system in a simulated
  12 hr integration after subtraction of the shock signal.\ The bulk ICM
  is modeled as an elliptical isothermal $\beta$ model with $R_{core}
  = (150, 250) \, {\rm kpc}$, $\beta = 0.7$, and $y_o = 3 \times
  10^{-5}$ at $z=0.7$, characteristic of disturbed, merging systems.}
\label{fig:szshock}
\end{center}
\end{figure}

ALMA Band 1 will have a considerably higher sensitivity for these
observations than the JVLA, owing to an order of magnitude
higher surface brightness sensitivity, or ALMA Band 3, owing to
lower system temperatures and larger primary beam. We simulated
Band 1 and Band 3 observations (50 12-m's and ACA) covering the
virial region ($D\sim 5'$) of a moderately massive SZE cluster with a
merger shock (Figure~\ref{fig:szshock}).  We considered a hypothetical
project aiming to detect a feature with a Compton $y = 10^{-4}$ -- a
characteristic of strong shocks in major mergers -- over the virial
radius of a cluster, with a characteristic feature size of
$5''-20''$. The required flux density sensitivity is similar in both
cases after allowing for resolution effects, about $8-9 \, {\rm \mu
Jy}$ RMS ($1 \sigma$) in both instances.  We find that a clear
detection is achieved in only $1.5$ hours of Band 1 observing, but
nearly $40$~hours are required at Band 3. The ACA Band 1 measurement
of the bulk ICM signature (a 12 hr observation is needed for good SNR)
is also shown, tapered to a $45''$ FWHM beam. Yamada et al.\ (2012)
find similar results in a detailed study of SZE imaging with ALMA and
the ACA at $\lambda \approx 1 \, {\rm cm}$.

\subsection{Line Observations with ALMA Band 1}

\subsubsection{Fine Structure of Chemical Differentiation in Cloud Cores}

Previous single-dish millimeter molecular-line observations have found that molecular 
distributions differ significantly between individual dark cloud cores.  A widely accepted 
interpretation of this chemical differentiation is that there exists non-equilibrium 
gas-phase chemical evolution through ion-molecule reactions within dark cloud cores.
Younger cores are rich  in ``early-type" carbon-chain molecules such as CCS and HC$_3$N, 
while more evolved cores, closer to protostellar formation via gravitational collapse, are 
rich in ``late-type" molecules such as NH$_3$ and SO (Suzuki et al.~1992). 
Recent high-resolution millimeter-line observations, however, have revealed that
there are even finer variations of molecular distributions within cores down to $\sim$3000 AU
scales, and that these fine-scale chemical fluctuations cannot be explained by the simple scenario of chemical evolution of cores (Takakuwa et al.\ 2003, Buckle et al.\ 2006).  The suggestion
is that there is first molecular depletion onto grain surfaces in these regions and then 
subsequent reaction and desorption
of molecules back to the gas phase through clump-to-clump collisions or energy injection from 
newly formed protostars (e.g., Buckle et al.\ 2006).  The molecules that can differentiate between regions with 
``early--type" chemistry, before any collapse of a protostellar object, and the ``late--type" chemistry that becomes 
apparent after the formation of a protostellar core, have their ground-state (strongest) transitions in ALMA Band 
1. These heavy saturated organic molecules can only be formed on the surfaces of dust grains, and so their 
appearance in the interstellar medium signals the presence of a central heating source, likely a protostar. ALMA 
Band 1 will provide the most sensitive test of when a central heating source turns on, because it will have the 
resolution and sensitivity to detect the presence of these saturated complex molecules within a dense core of 
more diffuse, unprocessed gas.

Other recent work (see Garrod, Weaver \& Herbst 2008 and references therein)
has shown some surprising detections of saturated complex organic molecules
around apparently quiescent dust cores, consistent with model predictions for the
``warm-up" chemistry expected when a core is undergoing gravitational collapse
and forming an internal heating source.  According to models, a later stage in this
sequence occurs when complex saturated molecules produced on grain surfaces
react as the gas warms up, producing ``hot core" chemistry, with even more
complex products.

To probe the smallest length scales of chemical variations in cloud cores and to clarify the
relationship among different molecular distributions (in conjunction with chemical models) 
requires both ALMA's excellent spatial resolution and its ability to recover the large-scale 
structure of the cloud through observations with the ACA.  Table~\ref{table:mol} lists some
molecular transitions needed for the chemical studies within these clouds that are observable
over 35-52 GHz.

\begin{table*}[h]

\caption{Molecular Transitions between 35 GHz and 52 GHz}
\label{table:mol}

\begin{center}
\begin{tabular}{lcr}
\hline
SO & 2$_3$--2$_2$& 36.202040 GHz\\
HC$_3$N & 4--3& 36.392332 GHz\\
HCS$^+$ &1--0 & 42.674205 GHz\\
SiO & 1--0 & 43.42376 GHz\\
HC$_5$N & 17--16 & 45.264721 GHz\\
CCS & 4$_3$--3$_2$& 45.379033 GHz\\ 
HC$_3$N & 5--4 & 45.490316 GHz\\
CCCS & 8--7& 46.245621 GHz\\
C$_3$H$_2$ & 2$_{1,1}$--2$_{0,2}$& 46.755621 GHz\\
C$^{34}$S & 1--0& 48.206956 GHz\\
CH$_3$OH & 1$_0$--0$_0$& 48.372467 GHz\\
CS & 1--0 & 48.99096 GHz\\
HDO & 3$_{2,1}$--3$_{2,2}$ & 50.23630 GHz\\
HC$_5$N & 19--18 & 50.58982 GHz\\
DC$_{3}$N & 6--5 & 50.65860 GHz\\
O$_{2}$ & N=35-35, J=35-34 & 50.98773 GHz\\
CH$_{3}$CHO & 1(1,1)-0(0,0) & 51.37391 GHz\\
NH$_{2}$D & 1(1,0)--1(1,1) & 51.47845 GHz\\
CH$_{2}$CHCHO & 1$_{11}$--0$_{00}$ & 51.59607 GHz\\
C$_3$H$_2$ & 1$_{1,1}$--0$_{0,0}$& 51.841418 GHz\\
\hline
\end{tabular}
\end{center}
\end{table*}

\subsubsection{Complex Carbon Chain Molecules}

ALMA Band 1 will provide a unique opportunity to search for new complex
organic molecules, including the amino acids and sugars from which life on
Earth may have originally evolved.  In addition, these complex molecules
provide a powerful tool for understanding star formation and the processes
surrounding it.

There are several reasons why Band 1 is the best place to search for
complex molecules. First, the heavier a molecule, the lower will be its 
rotational transition frequencies. The many abundant lighter molecules
(e.g., CO, HCN, CN) have their lowest transitions in Band 3, and so do
not appear at all in Band 1. Therefore, Band 1 does not suffer from
contamination from these common molecules, and so line confusion
is much less of a problem.  Second, the system temperatures in Band 1
are significantly lower than in higher bands, giving extra sensitivity to
detect weak transitions from less abundant complex molecules, such
as glycolaldehyde, the simple sugar known to exist in the interstellar
medium.  Table~\ref{table:ccm} lists some complex carbon-chain
molecules whose transitions have been already detected in the ISM.
Note that searches for complex molecules can be made with Band 1
using emission lines, but also using lines in absorption against bright
background objects like, e.g., young stars or quasars.

There is now a significant body of evidence to suggest that complex
biological molecules, such as amino acids and sugars needed for
evolution of life on Earth, evolved in the interstellar medium (e.g.,
see Holtom et al.\ 2005; Hunt-Cunningham \& Jones 2004; Bailey 
et al.\ 1998).  Band 1 will be one of the best instruments in the world
to test this hypothesis observationally. 

\begin{table*}[h]

\caption{Some detected ISM complex carbon chain molecules}
\label{table:ccm}

\begin{center}

\begin{tabular}{lr}

\hline

CH$_2$CHCN & propenitrile\\

CH$_2$CNH & ketenimine\\

CH$_3$C$_4$H & methyldiacetylene\\

CH$_3$CCCN & methyl cyanoacetylene\\

CH$_3$CH$_2$CN & ethyl cyanide\\

CH$_3$CHO & acetaldehyde\\

CH$_3$CONH$_2$ & acetamide\\

CH$_3$OCH$_3$ & ethyl butyl ether\\

CH$_3$OCHO & methyl formate\\

C$_6$H$^-$ & hexatriyne anion\\

C$_8$H & octatetraynyl\\

H$_2$CCCC & cumulene carbene\\

HCCCNH$^+$ & \nodata \\

\hline
\end{tabular}
\end{center}
\end{table*}

\subsubsection{Radio Recombination Lines}

Radio recombination lines (RRLs) are powerful, extinction-free diagnostics of the ionized gas in young, star-forming regions.  Dozens of galaxies have been detected in RRLs so far, and improved sensitivity ALMA Band 1 and the JVLA will open new capabilities in galactic and extragalactic RRL studies (e.g., Peters et al. 2012).  ALMA and the JVLA will also each have a significant impact on the study of photoevaporating protoplanetary disks using RRLs (Pascucci et al.\ 2012).  RRLs do not suffer from dust obscuration and thus provide a powerful method for studying the kinematics, structure, and physical properties of ionized gas such as the ionizing photon flux, density, filling factor, and electron temperature in the nuclear regions of dusty galaxies. A comparison of the predicted strengths of NIR recombination lines (e.g., Br$\alpha$ and Br$\gamma$ lines) based on RRLs with actual observations also leads to revised estimates of the dust extinction, which are essential for probing the heavily dust-enshrouded nuclei of galaxies such as Arp 220 (e.g., Kepley et al.\ 2011). 

RRLs are observed over a wide range of frequency, from a few GHz up to a few 100 GHz (e.g., Puxley et al.\ 1997).  It is important to observe RRLs at multiple, widely spaced frequencies because they trace different parts of the ionized medium (e.g., Anantharamaiah et al.\ 2000).  The observed strength of the millimeter-wavelength RRLs indicates the presence of a higher density ($n_e  \approx 10^5$ cm$^3$) ionized gas component (e.g., Zhao et al.\ 1997) and leads to information about star formation at recent epochs ($t \approx 10^5$ yr).  On the other hand, the lower frequency lines provide constraints on the amount of low density ($n_e \approx 10^3$ cm$^3$) ionized gas.  Therefore, adding a Band 1 receiver suite to ALMA is useful for studying the moderately dense ionized medium through observations of lines such as H53$\alpha$ (43.3094 GHz).  Accurate measurements of both the RRL and continuum emission are necessary to model the RRL emission, and ALMA gives an ideal combination of the large instantaneous band width and high spectral resolution.  To make an accurate determination of the free-free component, multi-wavelengths continuum measurements are also essential, so adding a Band 1 receiver suite is also beneficial for a better modelling of RRLs.

\subsubsection{Maser Science}

Masers (Microwave Amplifications by Stimulated Emission of Radiation) frequently occur in regions of active star formation, from molecular transitions whose populations are either radiatively or collisionally inverted. A photon emitted from this material will interact with other excited molecules along its path, stimulating further emission of identical photons. This process leads to the creation of a highly directional beam that has sufficient intensity to be detected at very large distances.

Masers are observed from a variety of molecular and atomic species and each serves as a signpost for a specific phenomenon, a property which renders masers powerful astrophysical tools (Menten 2007). More precisely, masers are formed under specific conditions, and the detection of maser emission therefore suggests that physical conditions (e.g., temperature, density, and molecular abundance) in the region where the maser forms lie within a defined range (c.f., Cohen 1995, Ellingsen 2004, and references therein). Therefore, interferometric blind and targeted surveys of maser species can lead to the detection of objects at interesting evolutionary phases (Ellingsen 2007).

\begin{table*}[h]
\caption{ALMA bands with known maser lines (Menten 2007)}
\label{table:maser1}
\newcommand{\m}{\hphantom{$-$}}
\newcommand{\cc}[1]{\multicolumn{1}{c}{#1}}
\renewcommand{\tabcolsep}{2.0pc} 
\renewcommand{\arraystretch}{1.2} 
\begin{center}
\begin{tabular}{@{}lclclcl}
\hline
 Species  &  ALMA Bands\\
\hline
 \m H$_2$O & \m B3, B5, B6, B7, B8, B9\\
 \m CH$_3$OH & \m B1, B3, B4, B6\\
 \m SiO& \m B1, B2, B3, B4, B5, B6, B7\\
 \m HCN & \m B3, B4, B6, B7, B9\\
 \hline
\end{tabular}\\[0.1pt]\
\end{center}
\end{table*}

Theoretical models of masers strongly depend on physical conditions as well as the geometry of the maser source. A successful model should be able to reproduce observational characteristics of observed maser lines but also to predict new maser transitions (e.g., the models of Sobolev 1997 for Class II methanol masers and Neufeld 1991 for water masers). In that respect, interferometry is essential for the successful search of candidate lines and confirmation of their maser nature.  ALMA, in particular, will resolve closely spaced maser spots and help further establish precise models of masing sources by determining if the detected maser signals are associated with thermal emission (Sobolev 1999), which is essential for improving theoretical models. With Band 1, ALMA will cover a wider frequency range, making it ideal for multi-transition observations of various maser species across the millimeter and  submillimeter windows. Examples of species with observed maser radiation in the different ALMA bands are given in Table \ref{table:maser1}, while Tables \ref{table:maser2} \& \ref{table:maser3} list SiO and methanol maser transitions that have been observed or predicted to be within Band 1.  

Maser radiation can be linearly or circularly polarized depending on the magnetic properties of the molecule. Polarimetric studies of maser radiation with interferometers can therefore yield information on the morphology of the magnetic field threading the region on small scales, with the plane-of-sky and line-of-sight components of the field being probed using linear and circular polarization measurements, respectively (e.g., see Harvey-Smith 2008, Vlemmings 2006). Polarization data are essential for improving on the theory of maser polarization first introduced by Goldreich (1973a), which applies to a linear maser region, a constant magnetic field, the simplest energy states for a masing transition, and asymptotic limits. Observations at higher spatial resolution are needed to verify and improve on more realistic and extensive models (Watson 2008).

\begin{table*}[h]
\caption{Observed SiO maser lines in the Band 1 of ALMA (Menten 2007).}
\label{table:maser2}
\newcommand{\m}{\hphantom{$-$}}
\newcommand{\cc}[1]{\multicolumn{1}{c}{#1}}
\renewcommand{\tabcolsep}{2.0pc} 
\renewcommand{\arraystretch}{1.2} 
\begin{center}
\begin{tabular}{@{}lclclcl}
\hline
 Transitions  &  Frequency (GHz)\\
\hline
\m v=0 (J= 1 $\to$ 0) & \m 42.373359\\
  \m v=3 (J= 1 $\to$ 0) & \m 42.519373\\
\m v=2 (J= 1 $\to$ 0) & \m 42.820582\\
\m v=0 (J= 1 $\to$ 0) & \m 42.879916\\
 \m  v=1 (J= 1 $\to$ 0) & \m 43.122079 \\
\m v=0 (J= 1 $\to$ 0) & \m 43.423585\\
 \hline
\end{tabular}\\[0.1pt]\
\end{center}
\end{table*}

\begin{table*}[htb]
\caption{Observed (Menten 2007) and predicted (designated with a star, Cragg et al. 2005) methanol maser lines in Band 1}
\label{table:maser3}
\newcommand{\m}{\hphantom{$-$}}
\newcommand{\cc}[1]{\multicolumn{1}{c}{#1}}
\renewcommand{\tabcolsep}{2.0pc} 
\renewcommand{\arraystretch}{1.2} 
\begin{center}
\begin{tabular}{@{}lclclcl}
\hline
 Transitions  &  Frequency (GHz)\\
\hline
 \m  4(-1) $\to$ 3(0)E & \m 36.1693 \\
 \m 7(-2) $\to$ 8(-1)E & \m 37.7037\\
 \m 6(2) $\to$ 5(3)A$^+$ & \m 38.2933\\
\m 6(2) $\to$ 5(3)A$^-$ & \m 38.4527\\
\m 7(0) $\to$ 6(1)A$^+$ & \m 44.0694\\
\m 2(0) $\to$ 3(1)E  $^*$ & \m 44.9558\\
\m 9(3) $\to$ 10(2)E $^{*}$ & \m 45.8436\\
\hline
\end{tabular}\\[0.1pt]\
\end{center}
\end{table*}

\newpage

\subsubsection{Magnetic Field Strengths from Zeeman Measurements}

Magnetic fields are believed to play a crucial role in the star
formation process.  Various theoretical and numerical studies explain
how magnetic fields can account for the support of clouds against
self-gravity, the formation of cloud cores, the persistence of
supersonic line widths, and the low specific angular momentum of cloud
cores and stars (McKee \& Ostriker 2007).  The Òstandard
modelÓ  suggests that the initial mass-to-(magnetic) flux ratio,
M/$\Phi_{init}$, is the key parameter governing the fate of molecular
cores.  Namely, if M/$\Phi_{init}$ of a core is greater than the critical
value, the core will collapse and form stars on short time scales, but
for cores with M/$\Phi_{init}$ smaller than the critical value the
process of 
ambipolar diffusion will take a long time to reduce the magnetic
pressure (Mouschovias \& Spitzer 1976; Shu et al.\ 1987).  On the
other hand, recent MHD simulations suggest that turbulence can control
the formation of clouds and cores and in such cases the mass-to-flux
ratio in the center of a collapsing core will be larger than that in
its envelope, the opposite of the ambipolar diffusion results (Dib et
al.\ 2007).  Therefore, measuring the magnetic field strengths and the
mass-to-flux ratios in the core and envelope provide a critical test
for star formation theories.

Despite its central importance, the magnetic field is the most poorly
measured parameter in the star formation process.  The main problem is
that magnetic fields can be measured only via polarized radiation,
which requires extremely high sensitivity for detections.  As a
result, the observed data on magnetic fields is 
sparse compared with those related to densities,
temperatures, and kinematics in star-forming cores.  The large
collecting area of ALMA provides the best opportunity to resolve the
sensitivity problem for magnetic field measurements.

The key to determining mass-to-flux ratios is 
the measurement of the strength of magnetic fields.
This measurement can be made {\it directly} through detection
of the Zeeman effect in spectral lines.
Observations of Zeeman splitting involve detecting the small
difference between left and right circular polarizations, which is
generally very small in interstellar conditions 
(with the exception of masers).
Successful non-maser detections of 
the Zeeman effect in molecular clouds
have only been carried out with HI, OH, and CN lines 
because these species have the
largest Zeeman splitting factors ($\sim$2 -- 3.3 Hz/$\mu$G) among all
molecular lines (Crutcher et al.\ 1996, 1999; Falgarone et al.\ 2008).
Thermal HI and OH lines, however, probe relatively low-density gas
($n$(H) $< 10^4$ cm$^{-3}$).  Also, CN detections are difficult; 
Crutcher (2012) described only 8 CN Zeeman detections towards
14 positions observed with significant sensitivity.

ALMA Band 1 receivers provide the opportunity to detect the Zeeman
effect from the CCS 4$_{3}$--3$_{2}$ line at 45.37903 GHz and hence
greatly advance our understanding in star formation.  CCS has been
widely recognized as being present only very early in the star-forming
process through chemical models (Aikawa et al.\ 2001, 2005) and
observations (Suzuki et al.\ 1992; Lai \& Crutcher 2000). Therefore the
mass-to-flux ratio derived from the CCS Zeeman measurements will be
very close to the initial values before the onset of gravitational
collapse.  CCS 4$_{3}$--3$_{2}$ also has a relatively large Zeeman
splitting factor ($\sim$ 0.6 Hz/$\mu$G; Shinnaga \& Yamamoto 2000)
compared to most molecules.  ALMA's antennas and site will be excellent
at these ``long'' wavelengths, providing the stability and accuracy needed
for such sensitive polarization work. The linearly polarized detectors on
ALMA's antennas will also be ideally suited to measurement of Stokes
V signatures from CCS.

\begin{figure} 
\includegraphics[scale=0.9]{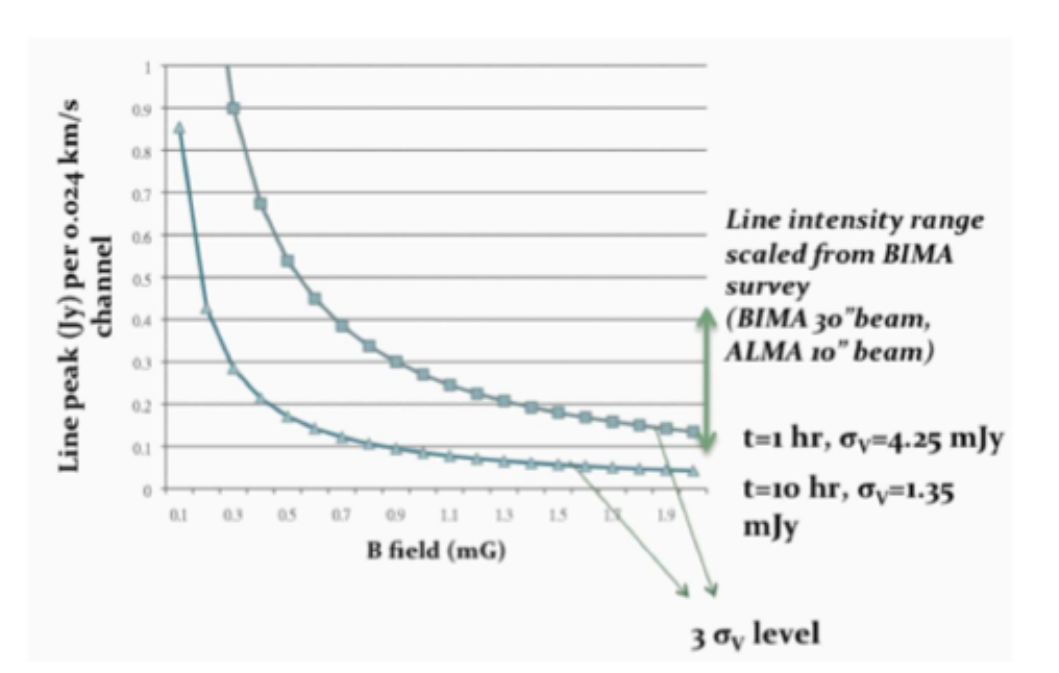}
\caption{The expected detection limits (3 $\sigma$) with integration time of 1 hr and 10 hr for a range of magnetic field strengths and CCS line intensity.}
\label{fig:ccs}
\end{figure}

Using the BIMA survey results from Lai \& Crutcher (2000), 
Figure~\ref{fig:ccs}
demonstrates that detections of CCS Zeeman effects can be achieved 
if the ALMA specifications for Band 1 receivers are met.
Zeeman effect
detection depends on two factors: the magnetic field strength and the line
intensity.  The two lines in Fig.~\ref{fig:ccs} show the 3 $\sigma$ detection
limits for Stokes V spectrum
with channel width of 0.024 km s$^{-1}$ and 1 hr or 10 hr integration time
for a range of magnetic field strengths and line intensities.  The channel
width is chosen to have at least 6 channels across the FWHM of the total
intensity spectrum (Stokes I).  If we scale the line intensity from Lai \&
Crutcher (2000) assuming the intensity distribution is uniform within
the 30$\arcsec$ BIMA beam, the expected line intensity would be around
0.1-0.4 Jy for ALMA observations with 10$\arcsec$ beam. Therefore, 
Fig.~\ref{fig:ccs} shows that for the magnetic fields of 0.2-1 mG (typical
values estimated from the application of the Chandrasehkar-Fermi method
to dust polarimetry in dense cores), we can detect the CCS Zeeman effect
with reasonable on-source integration time (less than 10 hr).

Note that the SiO v=1, J=1--0 transition at 43.12 GHz could be also used to
probe magnetic fields using the Zeeman effect, under certain circumstances.
Though its Zeeman splitting factor is lower than that of the above CCS lines,
the Zeeman effect may be detectible in situations where the SiO line is
extraordinarily bright, e.g., as a maser (see McIntosh, Predmore \& Patel
1994).  (Non-Zeeman interpretations of circularly polarized SiO emission
have also been advanced; see Weibe \& Watson 1998).

In summary, ALMA Band 1 receivers will provide the 
opportunity to measure the initial mass-to-flux ratio of molecular
cores through the detection of the Zeeman effect, which cannot be done
with any other instruments in the foreseeable future.  The results
will allow us to test realistically the expectations from theoretical
and numerical models for the first time.

\subsubsection{Jets from Young Stars}

\begin{figure} 
\epsscale{0.8}
\includegraphics[scale=0.8]{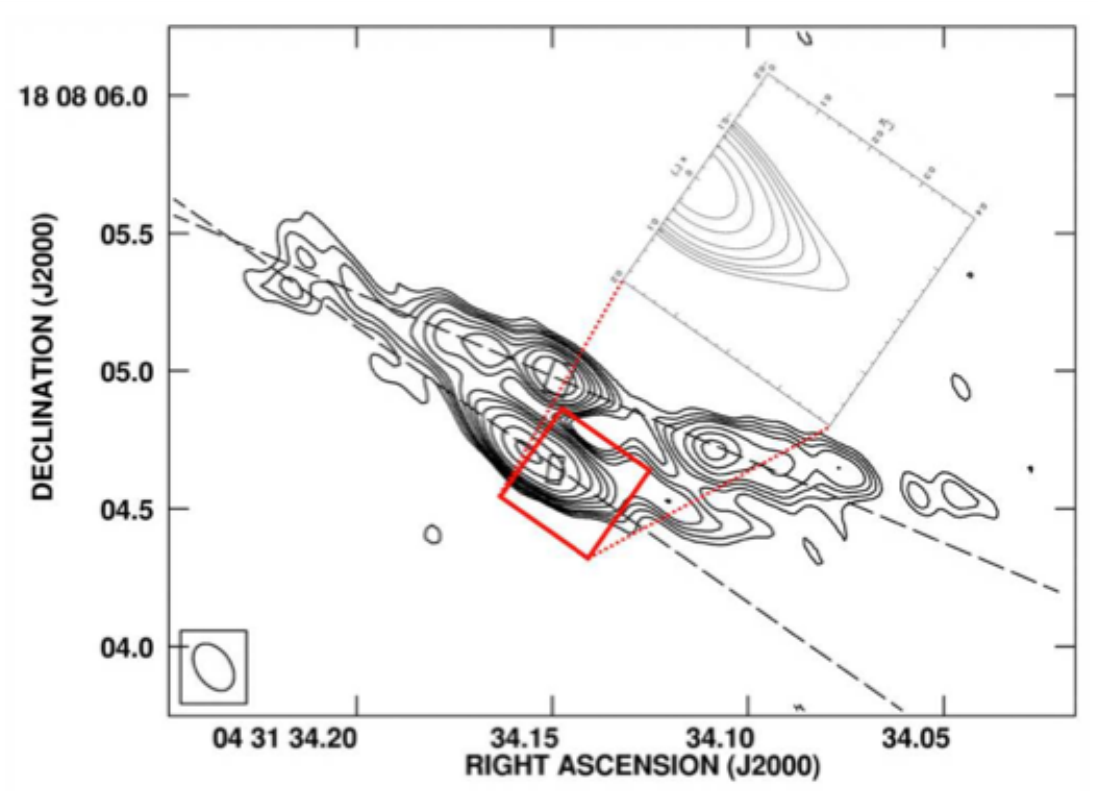}
\caption{{\bf Background}: The VLA+Pie Town continuum image of L1551 IRS 5 at 3.5 cm obtained by Rodriguez et al.\ (2003) in their Figure 1. The size of the beam (0.18 X 0.12\arcsec; P.A. =  35\degree) is shown in the bottom left-hand corner. Black rectangles mark the positions and deconvolved dimensions of the 7 mm compact protoplanetary disks. The dashed lines indicate the position angles of the jet cores. {\bf Inset}: map of the south jet from the X-Wind model convolved with the beam and plotted with the same contour levels from Figure 4 of Shang et al.\ (2004).
}
\label{fig:shang}
\end{figure}

Radio continuum emission is observed from the jets and winds of young stellar objects and is due to the interaction of free electrons, i.e., ``free-free emission." The radio images appear elongated and jet-like and are usually located near the base of large optical Herbig-Haro flows (Reipurth \& Bally 2000). These regions usually have only sub-arcsecond sizes, indicating the youth of the emitting material and the short dynamical times involved.  The emitted flux is usually weak, with a flat to positive spectral index with increasing frequency, and it can be obscured by the stronger thermal emission from dust grains at higher frequencies (e.g.,  Anglada 1995).  Multi-wavelength studies of the brightest radio jets at centimeter wavelengths trace either earlier and stronger sources or more massive systems. The triple system L1551-IRS 5, one of the most studied low-mass systems (Rodriguez et al.\ 1998, 2003; Lim \& Takakuwa 2006), is illustrative of the sub-arcsecond scales required (Figure~\ref{fig:shang}).

Ground-based, interferometric studies of radio jets provide the best opportunity to resolve the finest scales of the underlying source, comparable or better than optical studies of jets by the Hubble Space Telescope. Such fine-detail images can provide the ability to differentiate between theoretical ideas about the nature of these jets; i.e., the launch region, the collimation process, and the structure of the inner disks. Modeling efforts with the radio continuum emission presented in Shang et al.\ (2004) demonstrate one such possibility in constraining theoretical parameters using earlier millimeter and centimeter interferometers (Figure~\ref{fig:shang}).  Band 1 observations will discriminate between competing jet launch theories tied to the disk location of the launch point by achieving better than 0.1$^{\prime\prime}$ angular resolution.

The high sensitivity of  Band 1 on ALMA will also allow detection of radio emission from less luminous sources. ALMA will thus have the potential to discover a significant number of new radio jets, providing a catalog from which evolutionary changes in the physical properties can be deduced. As well, multi-epoch  surveys will be able to  follow the evolution of the freshly ejected material down to a few AU from the driving sources through movies. The 35-52 GHz frequency range of Band 1 will show contributions to the observed emission from both the ionized component of the jet and the thermal emission from the dust. Together with lower frequency JVLA observations and detailed theoretical modelling, a complete understanding of properties of the spectral energy distribution (SED) from the ionized inner regions of young stellar jets will be uncovered.

\subsubsection{Molecular Outflows from Young Stars}

The Sub-millimeter Array (SMA) has proven to be a successful instrument for the study of the youngest molecular outflows and jets from the most deeply embedded sources (e.g. Hirano et al.\ 2006; Palau et al.\ 2006; Lee et al.\ 2007a,b, 2008, 2009). The detection of excitation from rotational transitions of the molecule SiO up to levels $J=$ 8--7 and CO up to $J=$ 3--2 have uniquely identified a molecular high-velocity jet-like component located within the outflow shell.  This component displays similarities to the optical forbidden line jets observed in T-Tauri stars (Hirano et al.\ 2006; Palau et al.\ 2006; Codella et al.\ 2007; Cabrit et al.\ 2007). These observations have provided a new probe of how jets are launched and collimated during the earliest protostellar phase. 

One unique opportunity offered by Band 1 is observation of the $J=$ 1--0 transition of the SiO molecule at 43.424 GHz.  This transition has not yet been detected nor surveyed around even the brightest molecular outflows, except using single-dish telescopes (Haschick \& Ho 1990). One feature of this line that may be potentially distinct from the higher-J transitions of SiO is that it may be tracing the outer and more diffuse gas located on the outskirts of outflow shells that can be easily excited by shocks. Potential morphological and kinematic studies of the regions where the outflows interact with their own pre-natal clouds could be contrasted with other transitions using knowledge of their excitation conditions. 

\subsubsection{Co-Evolution of Star Formation and Active Galactic Nuclei}

Roughly half of the high-redshift objects detected in CO
line emission are believed to host an active galactic nucleus (AGN). 
Although they are selected based on their AGN properties, optically 
luminous high-redshift quasars exhibit many characteristics
indicative of ongoing star-formation, 
e.g., thermal emission from warm dust (Wang et al.\ 2008) or extended
UV continuum emission. Indeed, galaxies with AGNs in the local Universe 
reveal a strong correlation between the mass ($m$) in their supermassive
black hole (SMBH) and that in 
their stellar bulge (measured from the stellar velocity dispersion ($\sigma$); 
e.g., Kormendy \& Richstone 1995; Magorrian et al.\ 1998; Gebhardt et
al.\ 2000).  Such a correlation can be explained if the SMBH formed
coevally with the stellar bulge, implying that the  
luminous quasar activity signaling the formation of a sub-arcsecond SMBH at
high-redshift should be accompanied
by starburst activity. High spatial resolution observations of CO
line emission in high-redshift quasars 
can be used to infer the dynamical masses, which are found to be comparable
to the derived molecular gas + black hole masses, meaning that their
stellar component cannot contribute 
a large fraction of the total mass. There is mounting evidence that 
quasar host galaxies at redshifts $z$ = 4--6 have SMBH masses
up to an order of magnitude larger 
than those expected from their bulge masses and the local relation 
(Walter et al.\ 2004; Riechers et al.\  in prep.),
suggesting that the SMBH may have formed first.
The possible time evolution of the $m - \sigma$ relation is of
fundamental importance in studies of galaxy evolution, and 
this new finding needs to be made more statistically robust.
Future observations of high-redshift AGN with the Band 1 receivers
on ALMA would allow us to address this question through the study of 
\loj CO line emission in galaxies beyond redshifts $z \approx1.3$
(see \S4.2). 

\subsubsection{The Molecular Gas Content of Star-Forming Galaxies at $z \sim 2$}

While \loj CO line emission has only been detected in a few high-redshift objects, 
\hij CO line emission has been detected in more than sixty sources, most of which are classified as either submillimeter galaxies (SMGs) or far-infrared (FIR) luminous QSOs (see Carilli et al.\ 2011 for a review). Most of these studies have been conducted with sensitive interferometers and single-dish facilities operating in the 3~mm band (i.e., ALMA Band 3), which is sensitive to higher-J CO line transitions at high redshift, as is illustrated in Figure~\ref{fig:red2}. These lines generally trace warmer and denser gas, and so previous data may have led to a bias in our understanding of the molecular gas properties of high-redshift galaxies (e.g., Papadopoulos \& Ivison 2002). The addition of 35-50~GHz Band 1 receivers on ALMA will open up the possibility of studying the cold gas traced by the \loj transitions ({J}=2--1/1--0) in galaxies from moderate redshifts ($z \approx 1.3$) to those which existed when the Universe was reionized sometime before $z \ga 6$.

Although many previous studies of CO line emission in high-redshift
galaxies have focused on those starburst galaxies and AGN undergoing
episodes of extreme star formation (e.g., $\gg$100~M$_{\odot}$~yr$^{-1}$),
significant masses of molecular gas ($>10^{10}$~M$_{\odot}$) have been
discovered in more modest star-forming galaxies at $z = 1.5 - 2.0$
(Daddi et al.\ 2008). These ``BzK'' galaxies are selected for their
location in a B-$z$-K colour diagram (Daddi et al.\ 2004) and have
star-formation rates of $\sim$100~M$_{\odot}$~yr$^{-1}$ (Daddi et
al.\ 2007), while their number density is roughly a factor of 30
larger than that of the more extreme SMGs at similar
redshifts. Observations of \cotwo line emission in these BzK galaxies
reveal comparable masses of molecular gas to that of the SMGs, so
their star-formation efficiencies appear lower. The excitation
conditions of their molecular gas (temperature and density) are
similar to those of the Milky Way (Dannerbauer et al.\ 2008), as
indicated by the ``turnover'' in the CO line spectral energy
distribution occuring at the {J}=3--2 transition, i.e., lower than
that of the SMGs which typically occurs at the {J}=6--5 or {J}=5--4
transition (Weiss et al.\ 2005).  To develop a full spectral energy
distribution for the CO line excitation, observations of these
galaxies in the {J}=1-0 transition are needed with Band 1 receivers on
ALMA.  Such data will also provide a more robust
estimate of the total molecular gas mass, along with the spatial
resolution needed to constrain the gas kinematics, as has been
done for the SMGs (Tacconi et al.\ 2006).  Indeed, recent high-resolution
studies of CO $J$=1--0 from lensed Lyman Break galaxies (Riechers
et al.\ 2010) and unlensed BzK galaxies (Aravena et al., in prep.) 
have been made with the JVLA.  Also, CO $J$=1--0 emission has
been detected with the JVLA or GBT towards SMGs Ivison et al.\ 2010,
2011; Frayer et al.\ 2011; Riechers et al.\ 2011a,b).

\section{Other Considerations}
\label{sec:consid}
\subsection{Weather Considerations at the ALMA Site}

The ALMA site is exceptionally well-suited for Band 1 observing.  Even during the worst octile of weather,
the typical optical depth through the Band 1 Receiver range is less than 0.1.  Though other frequency ranges like
Band 3 can still use such weather, the addition of cloud cover and water droplets in the air will make 
still lower frequency observations more attractive.

As shown in the previous sections, the top science cases for Band 1 can stand shoulder-to-shoulder with the primary Level 0 goals of ALMA. Thus, while Band 1 observations will clearly benefit from a larger fraction of available observing conditions, the primary motivation for the enhancement is {\it not}\/ as a ``poor weather"
back-up receiver but rather the excellent science that can be achieved.

\subsection{ALMA and the Jansky Very Large Array}
The Jansky Very Large Array (JVLA) currently has observing capability over
the nominal Band 1 frequency range of 35--50 GHz, through its receivers in the 
K$_{\rm a}$-band (26.5--40 GHz) and Q-band (40--50 GHz).  Note, however, that
the JVLA cannot observe at 50-52 GHz, the extension proposed for ALMA Band 1.
Moreover, it is important to note that JVLA and ALMA are located at complementary
latitudes, the former at $+34\degree$ and the latter at $-23\degree$.  With both
telecsopes, the entire sky will be observable at high resolution at Band 1 frequencies.
(The southern hemisphere-based Australia Telescope Compact Array (ATCA) can
also observe some Band 1 frequencies but at much lower relative sensitivity than
ALMA or JVLA, and hence we do not consider it further.)  Here, we compare the
relative capabilities of ALMA and JVLA in Band 1.

Relative to ALMA, JVLA has fewer antennas (27 vs.\ 50) but these have larger 
surface areas (25-m diameter vs.\ 12-m), lower pointing accuracies (2-3\arcsec
vs.\ 0.6\arcsec) and lower aperture efficiencies at Band 1 (0.34--0.39 vs.\ 0.78).  
Combining these numbers (except pointing accuracy), the effective surface area
of JVLA is a factor of 1.02--1.17 times that of ALMA.  (Adding Band 1 to the ACA
antennas would minimize even this small difference.)  Due to its WIDAR correlator,
JVLA will have the same 8 GHz maximum bandwidth as ALMA.  
The WIDAR correlator can provide users with much higher spectral resolution 
than ALMA's correlator can, i.e., a maximum of below 1 Hz vs. 3.82 kHz.  
 
The JVLA is located at a lower altitude (2124 m) than ALMA (5000 m) and its 
local atmosphere has correspondingly larger typical precipitable water vapor.  
Weather statistics are needed for both the Plains of San Augustin and the 
Llano de Chajnantor to compare properly the availability of Band 1 observing 
conditions at both sites.  An important factor that also must be quantified 
is the stability of the atmosphere above both sites.

\begin{figure}
\begin{center}
\includegraphics[scale=0.6]{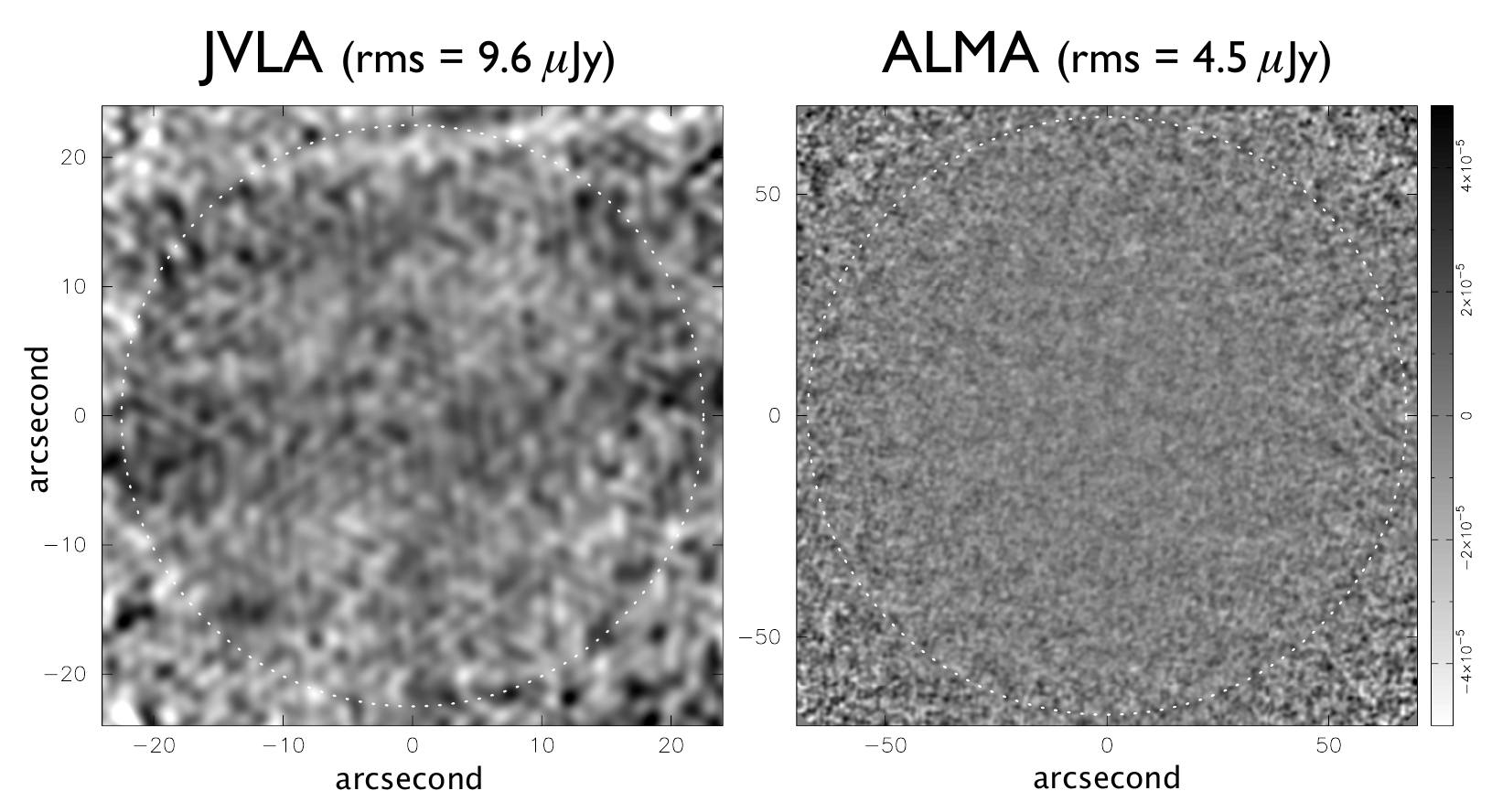}
\caption{Images from JVLA and ALMA observations simulated with CASA. 
The observations were set toward a ``blank" sky at 45~GHz with 8~GHz (continuum)
bandwidth, with JVLA in its D-configuration while ALMA in its \#12 configuration
provided in CASA.  Both array configurations give rise to a similar angular resolution of $\sim$ 1\farcs6
FWHM.  The white dotted circles denote the corresponding primary beam sizes.
There resulting 1 $\sigma$ rms noise levels after 2 hours of on-source integration
are 9.6 $\mu$Jy and 4.5 $\mu$Jy, respectively, for JVLA and ALMA, 
which are in general agreement with the estimated noise level shown in
Table~\ref{table:sensi_comp}.}
\label{fig:blank}
\end{center}
\end{figure}

Given differing antenna numbers, sizes, and baselines, the two observatories 
differ in various imaging metrics:
\begin{itemize}
	\item Comparing the face value ``single-field sensitivity" ($ND^{2}$; where $D$ is the
antenna diameter and $N$ is the number of antennas),  JVLA is over twice as ``sensitive" as ALMA (17000 vs.~7200).  
The ALMA Band 1 receiver specifications require receiver temperatures of 40--80 K over the Band, the 
same as the specification for the JVLA K$_{\rm a}$/Q-band receivers.  Assuming similar 
system temperatures for each set of receivers whenever local weather conditions 
are appropriate for Band 1 observing and factoring aperture efficiencies into account, the continuum sensitivities
for Band 1 are similar for both observatories\{footnote{Note also that for all sensitivity calculations
in this section we assume the original ALMA specifications for Band 1 receiver performance.}.
For example, a 1$ \sigma$ rms of $\sim$5 $\mu$Jy beam$^{-1}$ is expected at 35 GHz after 1
hour of integration at both observatories.  At higher frequencies (e.g., 45 GHz), however, the point
source sensitivity of ALMA is better than that of JVLA by a factor of $\sim$2.   Note also that the
JVLA sensitivities require the best weather while a relatively high PWV was chosen for ALMA in
these calculations.  Table~\ref{table:sensi_comp} provides comparison of JVLA and ALMA sensitivities
for point sources across the proposed Band~1 frequency coverage estimated using the JVLA and
ALMA sensitivity calculators ({\it https://science.nrao.edu/facilities/evla/calibration-and-tools/exposure}\/ and  
{\it http://almascience.eso.org/call-for-proposals/sensitivity-calculator}\/).  In addition, Figure~\ref{fig:blank} shows 
simulated ``blank-sky" observations at 45 GHz carried out with CASA, giving another
perspective on this comparison. (Note that additional parameters such as pointing accuracies
and phase stabilities have not been fully incorporated in these calculations.)

\begin{table*}[h]

\caption{Comparison of Point-Source Sensitivity between JVLA and ALMA}
\label{table:sensi_comp}
\begin{center}
\begin{tabular}{cccccc}
\hline
   \multicolumn{2}{c}{}                        & \multicolumn{2}{c}{JVLA}      & \multicolumn{2}{c}{ALMA} \\
   \multicolumn{2}{c}{no. of antennas}& \multicolumn{2}{c}{25}         & \multicolumn{2}{c}{50} \\
   \multicolumn{2}{c}{polarization}    & \multicolumn{2}{c}{dual}              & \multicolumn{2}{c}{dual} \\
   \multicolumn{2}{c}{weather}         & \multicolumn{2}{c}{winter}    & \multicolumn{2}{c}{auto (5.2mm) PWV} \\
   \multicolumn{2}{c}{source position} & \multicolumn{2}{c}{zenith}    & \multicolumn{2}{c}{zenith} \\
   \multicolumn{2}{c}{weighting}               & \multicolumn{2}{c}{natural}   & \multicolumn{2}{c}{natural} \\
   \multicolumn{2}{c}{on-source time}  & 60~s & 1~hrs & 60~s & 1~hrs \\
\hline
   \multicolumn{2}{c}{bandwidth}       & \multicolumn{2}{c}{1MHz}      & \multicolumn{2}{c}{1MHz} \\
   freq. & 35 GHz   & 3.2~mJy   &  0.41~mJy   &  3.0 mJy  & 0.38 mJy   \\
            & 40 GHz   & 3.6~mJy   &  0.47~mJy   &  3.1 mJy  & 0.40 mJy   \\
            & 45 GHz   & 5.1~mJy   &  0.68~mJy   &  3.6 mJy  & 0.47 mJy   \\
            & 50 GHz   & 25.5~mJy   &  3.29~mJy   & (not avail.)  & (not avail.)   \\
\hline
   \multicolumn{2}{c}{bandwidth}               & \multicolumn{2}{c}{8GHz}      & \multicolumn{2}{c}{8GHz} \\
    freq. & 40 GHz   & 50~$\mu$Jy   &  5.4~$\mu$Jy   & 35~$\mu$Jy  & 4.5~$\mu$Jy   \\
    freq. & 45 GHz   & 78~$\mu$Jy   &  10~$\mu$Jy   & 41~$\mu$Jy  & 5.3~$\mu$Jy   \\
\hline
\end{tabular}
\end{center}
\end{table*}

	\item  Comparing ``mosaic image sensitivity"  ($ND$), again on face value, JVLA and 
ALMA are similar (680 vs.~600).  JVLA's smaller number of baselines, however, yield a lower 
``image fidelity" ($N^3$) by a factor of $>6$ (20,000 vs.~130,000).  JVLA's larger dishes, however, 
mean that its minimum baselines are longer than ALMA's (35-m vs.~16-m, see 
Table~\ref{table:beam_comp} for a comparison), suggesting lower sensitivity to
low surface brightness emission.

\begin{table*}[h]

\caption{Comparison of angular scale coverage between JVLA and ALMA at 45 GHz}
\label{table:beam_comp}
\begin{center}
\begin{tabular}{ccccc}
\hline
			& \multicolumn{2}{c}{JVLA}	& \multicolumn{2}{c}{ALMA} \\
\hline
 Configuration	& A & D & most extended & most compact \\
 B$_{min}$ (km)	& 0.68	& 0.035	& 0.04	& 0.015	\\
 B$_{max}$ (km) 	& 36.4	& 1.03	& 16		& 0.15 	\\
 $\theta_{PRIMARY}$& 60	& 60		& 135	& 135	\\
 $\theta_{HFBW}$	& 0.043	& 1.5		& 0.08	& 9		\\
 $\theta_{LAS}$	& 1.2		& 32		& 35		& 93		\\
\hline
\end{tabular}
\end{center}
\end{table*}

	\item At present, JVLA has maximum baselines that are a factor of 2 larger than ALMA's 
(36.4~km vs.~15-18~km), suggesting the JVLA can produce images of resolution up to a factor
of 2 higher than ALMA can at the same frequency.  
\end{itemize}

In summary, at 35--50 GHz, the JVLA and ALMA are effectively the same
instrument at very complementary latitudes and with somewhat different strengths.
JVLA can go to lower frequencies than 35 GHz but ALMA Band 1 may include 50--52
GHz.  Weather conditions at both observatories suggest Band 1 observing conditions
are relatively rare for JVLA but should be common for ALMA, but the relative
fractions of available ``Band 1 time" need to be determined.  Also, the fraction
of time at ALMA where no other frequency but Band 1 can be observed needs
to be determined.

\newpage
\section{Summary}
\label{sec:conc}

The Band 1 receiver suite has been considered an essential part of ALMA from the earliest planning days. Even through the re-baselining exercise in 2001, the importance of Band 1 was emphasized. With the ALMA Development Plan taking shape, we have undertaken an updated review of the scientific opportunity at these longer wavelengths.  This document presents a set of compelling science cases over this frequency range.  The science cases reflect the new proposed range of Band 1, 35-50 GHz (nominal) with an extension up to 52 GHz, which was in fact chosen to optimize the science return from Band 1.  The science cases range from nearby stars and galaxies to the re-ionization edge of the Universe.  Two provide additional leverage on the present ALMA Level One Science Goals and are seen as particularly powerful motivations for building the Band 1 receiver suite: (1) detailing the evolution of grains in protoplanetary disks, as a complement to the gas kinematics, requires continuum observations out to $\sim 35\,$GHz ($\sim 9\,$mm); and (2) detecting CO 3 -- 2 spectral line emission from Galaxies like the Milky Way during the era of re-ionization, $6 < z < 10$
also requires Band 1 receiver coverage.


\newpage

\section{References}
\label{sec:refs}

\noindent
Acke, B., \& van den Ancker, M. E.\ 2004, A\&A, 426, 151

\noindent
Acke, B., et al.\ 2004, A\&A, 422, 621

\noindent
Adams, F. C., Emerson, J. P., \& Fuller, G. A.\ 1990, ApJ, 357, 606

\noindent
Anantharamaiah, K. R., Viallefond, F., Mohan, N. R., Goss, W. M., \& Zhao, J. H. 2000, ApJ, 537, 613

\noindent
Anglada, G.\ 1995, RMAACS, 1, 67

\noindent
Aikawa, Y., Ohashi, N., Inutsuka, S.-I., Herbst, E., \& Takakuwa, S.\ 2001, \apj, 552, 639

\noindent
Aikawa, Y., Herbst, E., Roberts, H., \& Caselli, P.\ 2005, \apj, 620, 330

%

\noindent
Andrews, S., \& Williams, J. P.\ 2007a, ApJ, 659, 705

\noindent
Andrews, S., \& Williams, J. P.\ 2007b, ApJ, 671, 1800


\noindent
Bailey, J., Chrysostomou, A., Hough, J. H., Gledhill, T. M., McCall, A., Clark, S., Menard, F., \& Tamura, M. 1998, Science, 281, 672

\noindent
Bartel, N.\ et al.\ 2002, ApJ, 581, 404

\noindent
Bartel, N., Bietenholz, M.F., Rupen, M.P., \& Dwarkadas, V.V.\ 2007, ApJ, 668, 924

\noindent
Bauer, F. E., et al.\ 2008, ApJ, 688, 1210

\noindent
Beckwith, S., Sargent, A. I., Chini, R. S., \& Guesten, R.\ 1990, AJ, 99, 924

\noindent
Beckwith, S., \& Sargent, A. I.\ 1991, ApJ, 381, 250

\noindent
Bertoldi, F.\ et al.\ 2003, A\&A, 409, L47

\noindent
Bietenholz, M. F., Bartel, N., \& Rupen, M. P.\ 2004, Science, 304, 1947

\noindent
Bietenholz, M. F., Bartel, N., \& Rupen, M. P.\ 2010, ApJ, 712, 1057

\noindent
Blain, A.~W., Smail, I., Ivison, R.~J., Kneib, J.-P., \& Frayer, D.~T.\ 2002, \physrep, 369, 111 

\noindent
Boley, A., et al. 2012, \apj, 750, L21

\noindent
Bonamente, M., Joy, M., LaRoque, S.~J., et al.\ 2008, \apj, 675, 106 

\noindent
Birkinshaw, M.\ 1999, Phys. Rep.\ 310, 97

\noindent
Buckle, J. V., et al.\ 2006, Faraday Discussions, 133, 63

\noindent
Bouwens, R. J., et al.\  2009, \apj, 690, 1764

\noindent
Boss, A.\  2005, ApJ, 629, 535

\noindent
Cabrit, S., et al.\ 2007, \aap, 468, L29

\noindent
Calvet, N., et al.\  2002, ApJ, 568, 1008

\noindent
Cappelluti, N., Predehl, P., B{\"o}hringer, H., et al.\ 2011, Memorie della Societa Astronomica Italiana 
Supplementi, 17, 159 

\noindent
Carilli, C. L., et al.\ 2007, ApJ, 666, L9

\noindent
Carilli, C. L., et al.\ 2008, Ap\&SS, 313, 307

\noindent
Carlstrom, J. E., Holder, G. P., \& Reese, E. D.\ 2002, \araa, 40, 643

\noindent
Casassus, S., et al.\ 2008, MNRAS, 391, 1075

\noindent
Chevalier, R.A.\ 1982, ApJ, 259, 85

\noindent
Cohen, R. J.\ 1995, Ap\&SS, 224, 55

\noindent
Codella, C., et al.\ 2007, \aap, 462, L53

\noindent
Cool, R.J., et al.\  2006, \aj, 132, 823


\noindent
Cragg, D. M., Sobolev, A. M., \& Godfrey, P. D.\ 2005, MNRAS, 360, 533

\noindent
Crutcher, R. M. 2012, ARA\&A, 50, 29

\noindent
Crutcher, R. M., Troland, T. H., Lazareff, B., \& Kazes, I.\ 1996, \apj, 456, 217

\noindent
Crutcher, R. M., Troland, T. H., Lazareff, B., Paubert, G., \& Kaz{\`e}s, I.\ 1999, ApJL, 514, 121

\noindent
Daddi, E., et al.\ 2004, ApJ, 617, 746 

\noindent
Daddi, E., et al.\ 2007, ApJ, 670, 156 

\noindent
Daddi, E., et al.\ 2008, ApJL, 673, L21 

\noindent
Dannerbauer, H., et al.\ 2009, ApJL, 698, 178 

\noindent
Dark Energy Survey Collaboration, The 2005, arXiv:astro-ph/0510346 

\noindent
Dib, S., Kim, J., V{\'a}zquez-Semadeni, E., Burkert, A., \& Shadmehri, M.\ 2007, \apj, 661, 262

\noindent
Dodds-Eden, K., et al.\ 2009, \apj, 698, 676

\noindent
Draine, B.\  2003, ARA\&A, 41, 241

\noindent
Draine, B.\  2006, ApJ, 636, 1114

\noindent
Draine, B., \& Anderson, N.\ 1985, ApJ, 292, 494 

\noindent
Draine, B., \& Lazarian, A.\ 1998, ApJ, 508, 157

\noindent
Dullemond, C. P., \& Dominik, C.\ 2005, A\&A, 434, 971

\noindent
Dunkley, J., et al.\ 2009, ApJS, 180, 306

\noindent
Dunkley, J., Hlozek,  R., Sievers, J., et al.\ 2011, \apj, 739, 52 

\noindent
Eckart, A., et al.\ 2006, \aa, 450, 535

\noindent
Ellingsen, S. P.\ et al.\  2007, IAUS, 242, 213

\noindent
Ellingsen, S. P.\  2004, IAUS, 221, 133

\noindent
Ettori, S., \& Fabian, A.~C.\ 2006, \mnras, 369, L42 

\noindent
Falgarone, E., Troland, T. H., Crutcher, R. M., \& Paubert, G.\ 2008, \aap, 487, 247

\noindent
Fan, X., et al.\ 2001, AJ, 122, 2833

\noindent
Fan, X., et al.\  2006a, \aj, 132, 117

\noindent
Fan, X., Carilli, C. L., \& Keating, B.\  2006b, \araa, 44, 415

\noindent
Fender, R. P., et al.\ 1999, ApJL, 519, 165

\noindent
Finkbeiner, D. P., Schlegel, D. J., Frank, C., \& Heiles, C.\ 2002, ApJ, 566, 898

\noindent
Frayer, D., et al.\ 2011, \apj, 726, L22

\noindent
Gaensler, B. M., et al.\ 2007, AIP Confer. Ser., ed.\ S.\ Immler \& R. McCray, 937, 86

\noindent
Garrod, R. T., Weaver, S. L. W., \& Herbst, E.\ 2008, \apj, 682, 283

\noindent
Gebhardt, K., et al.\ 2000, ApJ, 543, L5

\noindent
Geisbuesch, J., \& Hobson, M. P.\ 2007, \mnras, 382, 158

\noindent
Geisbuesch, J., Kneissl, R., \& Hobson, M. P.\ 2005, \mnras, 360, 41

\noindent
Ghez, A., et al. 2008, \apj, 689, 1044

\noindent
Gillessen, S., et al.\ 2009, \apj, 692, 1075

\noindent
Glikman, E., et al.\  2008, \aj, 136, 954

\noindent
Goldreich, P., Keeley, D. A., \& Kwan, J. Y.\ 1973, ApJ, 179, 111


\noindent
Greaves, J., Richards, A. M. S., Rice, W. K. M., \& Muxlow, T. W. B.\ 2008, MNRAS, 391, 74

\noindent
Guilloteau, S., et al. 20009, \aa, 529, 105

\noindent
Gunn, J.E., \& Peterson, B.A.\  1965, \apj, 142, 16331


\noindent
Hales, A. S., et al.\   2004, ApJ, 613, 977


\noindent
Haschick, A. D., \& Ho, P. T. P.\ 1990, \apj, 352

\noindent
Harvey-Smith, L., Soria-Ruiz, R., Duarte-Cabral, A., \& Cohen, R. J.\ 2008, MRAS, 384, 719 

\noindent
Helfand, D. J., Gotthelf, E. V., \& Halpern, J. P.\ 2001, ApJ, 556, 380

\noindent
Hincks, A. D., et al.\ 2009, arXiv:0907.0461

\noindent
Hirano, N., et al.\ 2006, ApJL, 636, 141

\noindent
Holtom, P. D., Bennett, C. J., Osamura, Y., Mason, N. J, \& Kaiser, R. I.\ 2005, \apj, 626, 940

\noindent
Hu, E.,M., Cowie, L. L., \& McMahon, R. G.\ 1998, ApJ, 502, L99

\noindent
Hunt-Cunningham, M. R., \& Jones, P. A.\ 2004, IAUS, 213, 159

\noindent
Isella, A, Natta, A., Wilner, D., Carpenter, J. M.. \& Testi, L.\ 2010, \apj, 725, 1735

\noindent
Ivison, R. J., et al.\ 2010, MNRAS, 404, 198

\noindent
Ivison, R. J., et al.\ 2011, MNRAS, 412, 1913

\noindent
Iye, M., et al.\ 2006, Nature, 443, 186

\noindent
Kepley, A. A., Chomiuk, L., Johnson, K. E., Goss, W. M., Balser, D. S., \& Pisano, D. J. 2011, ApJ, 739, L24

\noindent
Kitayama,T., Komatsu, E., Ota, N., Kuwabara, T., Suto, Y., Yoshikawa, K., Hattori, M., \& Matsuo, H. 2004, \pasj, 56, 17

\noindent
Koerding E. G., Fender R. P., \& Migliari S.\ 2006, MNRAS, 369, 1451

\noindent
Komatsu, E., Matsuo, H., Kitayama, T., Kawabe, R., Kuno, N., Schindler, S., \& Yoshikawa, K. 2001, \pasj, 53, 57

\noindent
Kormendy, J., \& Richstone, D.\ 1995, \araa, 33, 581

\noindent
Korngut, P.~M., Dicker,  S.~R., Reese, E.~D., et al.\ 2011, \apj, 734, 10 

\noindent
Lai, S.-P., \& Crutcher, R.M.\ 2000, \apjs, 128, 271

\noindent
Laki{\'c}evi{\'c}, M., Zarnado, G., van Loon, Th., Staveley-Smith, L., Potter, T., Ng, C.-Y., \& Gaensler, B.~M. 2012, A\&A, 541, L2 

\noindent
Lee, C.-F., et al.\  2007a, \apj, 659, 499

\noindent
Lee, C.-F., et al.\ 2007b, \apj, 670, 1188

\noindent
Lee, C.-F., et al.\ 2008, \apj, 685, 1026

\noindent
Lee, C.-F., et al.\ 2009, \apj, 699, 1584

\noindent
Lee, J., \& Komatsu, E.\ 2010, \apj, 718, 60 

\noindent
Leger, A., \& Puget, J. L.\ 1984, \aap, 137, L5

\noindent
Li, Y., et al.\ 2007, ApJ, 665, 187

\noindent
Li, Y., et al.\ 2008, ApJ, 678, 41

\noindent
Lim, J., \& Takakuwa, S.\ 2006, \apj, 653, 425

\noindent
Lonsdale, C. J., et al.\ 2006, ApJ, 647, 185

\noindent
Maccarone T. J., 2008, ASPC, 401, 191

\noindent
Maccarone T. J., Gallo E., Fender R. P.\ 2003, MNRAS, 345, L19

\noindent
Magorrian, J., et al.\ 1998, AJ, 115, 2285

\noindent
Maness, H.\ et al.\ 2008, ApJ, 686, L25

\noindent
Mannings, V., \& Emerson, J.\ 1994, MNRAS, 267, 361

\noindent
Markevtich, M., et al.\ 2000, \apj, 541, 542

\noindent
Markevitch, M., et al.\ 2002, \apj, 567, L27


\noindent
Markevitch, M., et al.\ 2009, arXiv:0902.3709

\noindent
Mason, B.~S., Dicker,  S.~R., Korngut, P.~M., et al.\ 2010, \apj, 716, 739 


\noindent
McIntosh, G. C., Read Predmore, C., \& Patel, N. A. 1994, \apj, 428, L29

\noindent
McKee, C. F. \& Ostriker, E. C. 2007, ARA\&A, 45, 565

\noindent
Melis, C., et al. 2011, \apj, 739, L7

\noindent
Menanteau, F., Hughes, J.~P., Sifon, C., et al.\ 2011, arXiv:1109.0953 

\noindent
Menten, K.\ 2007, IAUS, 242

\noindent
Merloni, A., et al. 2012, arXiv, 1209.3114

\noindent
Migliari S., Fender R. P., Rupen, M., Wachter, S., Jonker, P. G., Homan, J., \& van der Klis, M.\ 2004, MNRAS, 351, 186

\noindent
Morandi, A., Ettori, S., \& Moscardini, L.\ 2007, \mnras, 379, 518 


\noindent
Mortlock, D. J., et al.\ 2008, arXiv0810.4180

\noindent
Mouschovias, T.Ch.\  \& Spitzer, L., Jr.\ 1976, ApJ, 210, 326


\noindent
Narayanan, D., et al.\ 2008, ApJS, 174, 13

\noindent
Narayan, R., Ozel, F., \& Sironi, L. 2012, \apj, 757, L20

\noindent
Narayan R., \& Yi, I.\ 1994, ApJL, 428, 13

\noindent
Neufeld, D., \& Melnick, G.\ 1991, ApJ, 368, 215

\noindent
Ota, K., et al.\  2008, \apj, 677, 12

\noindent
Palau, A., et al.\ 2006,  ApJL, 636, 137

\noindent
Papadopoulos, P. P. \& Ivison, R. J.\ 2002, ApJ, 564, L9 

\noindent
Pascucci, I., Gorti, U., \& Hollenbach, D. 2012, ApJ, 751, L42

\noindent
Peters, T., Longmore, S. N., \& Dullemond, C. P. 2012, MNRAS, 425, 2352

\noindent
Pfrommer, C., En{\ss}lin, T.~A., \& Sarazin, C.~L.\ 2005, \aap, 430, 799 

\noindent
Plagge, T., et al.\ 2012, arXiv:1203.2175

\noindent
Planck Collaboration, Ade, P.~A.~R., Aghanim, N., et al.\ 2011, arXiv:1101.2024 

\noindent
Pinte, C., Menard, F., Duch\^{e}ne, G., \& Bastien, P.\ 2006, A\&A, 459, 797

\noindent
Pinte, C., et al.\  2009, A\&A, 498, 967

\noindent
Pollack, J., et al.\ 1996, Icarus, 124, 62

\noindent
Poole, G.~B., Babul, A., McCarthy, I.~G., et al.\ 2007, \mnras, 380, 437 

\noindent
Puxley, P. J., Mountain, C. M., Brand, P. W. J. L., Moore, T. J. T., \& Nakai, N. 1997, ApJ, 485, 143

\noindent
Rafikov, R.\ 2006, ApJ, 646, 288

\noindent
Reid, M. J. \& Brunthaler, A. 2004, \apj, 616, 872

\noindent
Reipurth, B., \& Bally, J.\ 2001, \araa, 39, 403

\noindent
Rephaeli, Y.\ 1995, \araa, 33, 541

\noindent
Rhoads J. E., et al.\ 2000,  ApJ, 545, L85

\noindent
Ricci, L., Testi, L., Natta, A., Neri, R., Cabrit, S., \& Herczeg, G. 2010, \aa, 512, 15

\noindent
Riechers, D., et al.\ 2009, \apj, 703, 1338

\noindent
Riechers, D., et al.\ 2010, \apj, 724, L153

\noindent
Riechers, D., et al.\ 2011a, \apj, 733, L11

\noindent
Riechers, D., et al.\ 2011b, \apj, 739, L31

\noindent
Rodmann, J., et al.\ 2006, A\&A, 446, 211

\noindent
Rodr{\'{\i}}guez, L. F., et al.\ 1998, \nat, 395, 355

\noindent
Rodr{\'{\i}}guez, L. F., et al.\ 2003, ApJL, 586, 137

\noindent
Sarazin, C.L.\ 1988, in: {\it X-ray emission  from clusters of galaxies}, Cam. Uni. Press

\noindent
Scaife, A. M. M., et al.\ 2009a, MNRAS, 400, 1394

\noindent
Scaife, A. M. M., et al.\ 2009b, MNRAS, submitted


\noindent
Sehgal, N., Trac, H., Acquaviva, V., et al.\ 2011, \apj, 732, 44 

\noindent
Shang, H., Lizano, S., Glassgold, A., \& Shu, F.\ 2004, ApJL, 612, 69

\noindent
Shinnaga, H., \& Yamamoto, S.\ 2000, \apj, 544, 330

\noindent
Shu, F. H., Adams, F. C., \& Lizano, S.\ 1987, \araa, 25, 23

\noindent
Sobolev, A. M., et al.\ 1999, in The Physics and Chemistry of the Interstellar Medium, ed. V. 
Ossenkopf et al.\ (GCA-Verlag: Herdecke), 299 

\noindent
Sobolev, A. M., Cragg, D. M., Godfrey, P. D.\ 1997, MNRAS, 288, L39


\noindent
Staniszewski, Z., et al.\ 2009, \apj, 701. 32

\noindent
Sutton, E. C., et al.\  2001, ApJ, 554, 173

\noindent
Sunyaev, R. A., \& Zel'dovich, Ya. B.\ 1972, Comm. Astrophys. Space  Phys. 4, 173

\noindent
Suzuki, H.,  et al.\ 1992, ApJ, 392, 551

\noindent
Tacconi L. J., et al.\ 2006, ApJ, 640, 228

\noindent
Takakuwa, S., Kamazaki, T., Saito, M., \& Hirano, N.\ 2003, ApJ, 584, 818

\noindent
Tananbaum H., Gursky H., Kellogg E., Giacconi R., \& Jones C.\ 1972, ApJ, 177, L5

\noindent
Taniguchi, Y., et al.\ 2005, PASJ, 57, 165

\noindent
Testi, L., Natta, A., Shepherd, D.S., \& Wilner, D.J.\ 2001, ApJ, 554, 1087

\noindent
Tudose, V., Fender, R. P., Linares, M., Maitra, D., \& van der Klis M.\ 2009, MNRAS, 400, 2111

\noindent
Vikhlinin, A., Kravtsov, A.~V., Burenin, R.~A., et al.\ 2009, \apj, 692, 1060 

\noindent
Vlemmings, W. H. T., Diamond, P. J., \& Imai, H.\  2006, IAUS, 234, 267

\noindent
Wagg, J., \& Kanekar, N.\ 2012, \apj, 751, L24

\noindent
Wagg, J., Kanekar, N., \& Carilli, C.L.\ 2009, ApJL, 697, L33

\noindent
Walter, F., et al.\ 2003, Nature, 424, 406

\noindent
Walter, F., et al.\  2004, \apjl, 615, L17

\noindent
Walter, F., et al.\  2009, \apjl, 691, L1

\noindent
Wang, R., et al.\  2008, AJ, 135, 1201

\noindent
Wang, R., et al.\  2010, \apj, 714, 699

\noindent
Wang, R., et al.\ 2011a, \aj, 142, 101

\noindent
Wang, R., et al.\ 2011b, \apj, 739, L34

\noindent
Watson, W. D.\ 2008, arXiv0811,1292W

\noindent
Weintraub, D. A., Sandell, G., \& Duncan, W. D.\ 1989, ApJ, 340, 69

\noindent
Weiss, A., Downes, D., Walter, F., \& Henkel, C.\ 2005, A\&A, 440, 45

\noindent
Welch, W. J., et al.\ 2004, in Bioastronomy 2002, eds R. Norris, F. Stootman (ASP, San Francisco), 213, 59

\noindent
Weiler, K. W., Panagia, N., Montes, M. J., \& Sramek, R. A. 2002, ARA\&A, 40, 387


\noindent
Weibe, D. S., \& Watson, W. D. 1998, ApJ, 503, L71

\noindent
Wik, D.~R., Sarazin, C.~L.,  Ricker, P.~M., \& Randall, S.~W.\ 2008, \apj, 680, 17 

\noindent
Williamson, R., Benson, B.~A., High, F.~W., et al.\ 2011, \apj, 738, 139 

\noindent
Willott, C. J., et al.\  2009, \aj, 137, 3541

\noindent
Wilner, D. J., Ho, P. T. P., Kastner, J. H., \& Rodriguez, L. F.\ 2000, ApJ, 534, 101

\noindent
Wootten, A.\  2007, IAUS, 242

\noindent
Wyatt, M. C., 2009, ARA\&A, 46, 339

\noindent
Yamada, K., et al.\ 2012, PASJ, in press.

\noindent
Yusef-Zadeh, F., et al. 2012, \apj, 757, L1

\noindent
Yusef-Zadeh, F., et al. 2006, \apj, 650, 189


\noindent
Zhao, J.-H., Anantharamaiah, K. R., Goss, W. M., \& Viallefond, F. 1997, ApJ, 482, 186

\newpage

\begin{figure}[t]
\begin{center}
\includegraphics[scale=0.8]{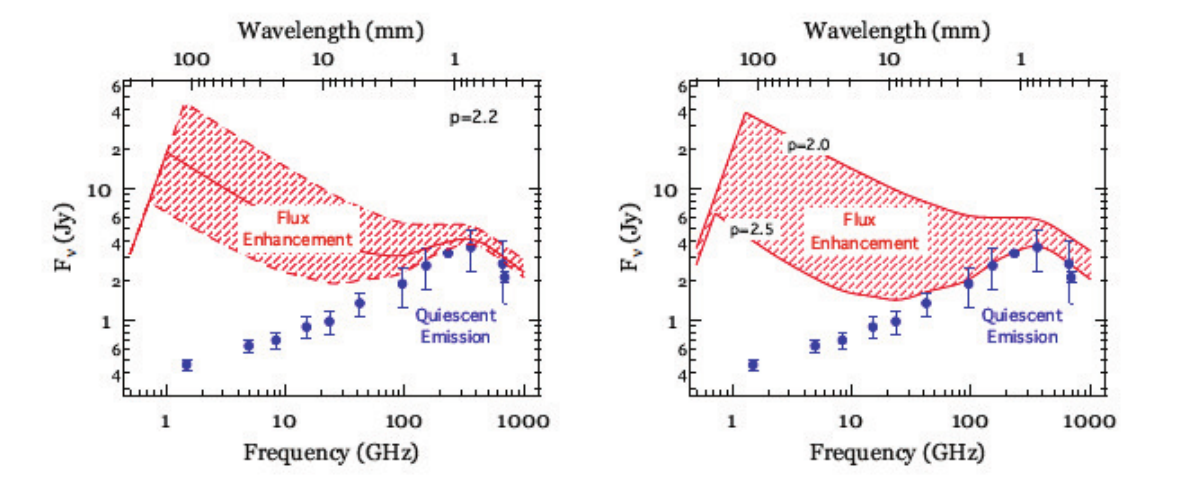}
\caption{
Radio emission as a function of frequency
expected from G2 cloud (red) when compared to
quiescent  emission from Sgr A*, as shown in blue (Narayan, Ozel, \& Sironi  2012). 
Left and right panels show predictions based on different assumptions
on the energy spectrum of nonthermal particles (p).
}
\label{fig:g2}
\end{center}
\end{figure}

\end{document}